\documentclass[12pt]{article}
\usepackage{amsmath, amssymb}
\usepackage{amsfonts}
\usepackage{graphicx,psfrag,epsf}
\usepackage{enumerate}
\usepackage{natbib, xcolor}
\usepackage{url} % not crucial - just used below for the URL 
\usepackage[noend]{algpseudocode}
\usepackage{algorithm}
\usepackage{booktabs}
\usepackage{multirow}
\usepackage{verbatim}
\usepackage{array}
\usepackage{caption}
\usepackage{setspace}
\usepackage{authblk}
\usepackage{rotating}
\usepackage{subcaption}
\usepackage[compact]{titlesec}
\newcommand{\blind}{0}

\newcommand{\btheta}{\mbox{\boldmath $\theta$}}
\newcommand{\bmu}{\mbox{\boldmath $\mu$}}

\newcommand{\bdelta}{\mbox{\boldmath $\delta$}}

\newcommand{\bSigma}{\mbox{\boldmath $\Sigma$}}

\algnewcommand\And{\textbf{and}}

\begin{document}

\def\spacingset#1{\renewcommand{\baselinestretch}%
{#1}\small\normalsize} \spacingset{1}

%%%%%%%%%%%%%%%%%%%%%%%%%%%%%%%%%%%%%%%%%%%%%%%%%%%%%%%%%%%%%%%%%%%%%%%%%%%%%%

\if0\blind
{

  \title{\bf Generalized Bayesian Additive Regression Trees for Restricted Mean Survival Time Inference}
\author[1]{Mahsa Ashouri}
\author[1]{Nicholas C. Henderson}
\affil[1]{{\small Department of Biostatistics, University of Michigan, Ann Arbor}}

\date{}
  \maketitle
} \fi

\if1\blind
{
  \bigskip
  \bigskip
  \bigskip
  \begin{center}
    {\LARGE\bf Parameter-Expanded ECME Algorithms for \\[2mm] Logistic and Penalized Logistic Regression}
\end{center}
  \medskip
} \fi

\bigskip
\begin{abstract}
Prediction methods for time-to-event outcomes often utilize 
survival models that rely on strong assumptions about noninformative censoring 
or on how individual-level covariates and survival functions are related. 
When the main interest is in predicting individual-level restricted mean survival times (RMST), reliance on such assumptions can lead
to poor predictive performance if these assumptions are not satisfied.
We propose a generalized Bayes framework that avoids full probability modeling
of all survival outcomes by using an RMST-targeted loss function
that depends on a collection of inverse probability of censoring weights (IPCW). 
In our generalized Bayes formulation, we utilize a flexible additive tree regression
model for the RMST function, and the posterior distribution of interest
is obtained through model-averaging IPCW-conditional
loss function-based pseudo-Bayesian posteriors. 
Because informative censoring can be captured by the IPCW-dependent
loss function, our approach only requires one to specify a
model for the censoring distribution, thereby obviating 
the need for complex joint modeling to handle informative censoring.
We evaluate the performance of our method through a series of simulations 
that compare it with several well-known survival machine learning methods, and 
we illustrate the application of our method using a multi-site cohort of breast cancer patients with clinical 
and genomic covariates.
\end{abstract}

\noindent%
{\it Keywords:} dependent censoring, ensemble methods, Gibbs posterior, inverse weighting, loss function, survival analysis
\vfill

\newpage
\spacingset{1.45} 

\section{Introduction}

The restricted mean survival time (RMST) is a measure that has emerged as a valuable quantity in survival analysis
due to its direct clinical interpretation, which holds regardless of the survival model used in an analysis (\citet{royston2013restricted}).
The RMST is defined as the expected survival duration up to a pre-specified truncation time $\tau$.
Changes in RMST are often cited as an alternative to hazard ratios in the context of treatment comparisons
and in survival regression modeling (\citet{pak2017, kloecker2020uses}). An advantage of using RMST is that differences in RMST
are model free in the sense that, unlike proportional hazards, the interpretation of change in RMST
holds regardless of whether or not a particular survival regression model holds. Moreover, 
constructing a prognostic model that directly targets the RMST as a function of baseline covariates
can be a useful tool because the interpretation of each estimated RMST is transparent and
does not require a proportional hazards assumption plus additional modeling of 
a baseline hazard function.

To estimate the RMST as a function of a patient's baseline covariates, there are a number of available approaches.
For instance, one could specify a full probability model for the survival outcomes, such as an accelerated
failure time (AFT) model where the RMST emerges as a function of other model parameters. Another possible option is to 
estimate RMST by using estimates of both the baseline hazard and regression parameters from a Cox proportional hazards model.
An alternative to these approaches that avoids full probability modeling of the survival outcomes is to 
plug in RMST-targeted ``pseudo-observations'' as the responses (e.g., \citet{andersen2004} and \citet{zhao2020}) and treat
these pseudo-observations as fully observed responses. 
One can then directly use more well-known regression or machine learning techniques to estimate RMST. 
Another attractive approach which also avoids full probability modeling is to use the inverse probability of censoring weights (IPCW; \citet{robins1993information}) as part of an objective function
that only uses the observed event times. IPCW-based approaches have been used by numerous authors in the context of
estimating regression models for RMST. For example, \citet{tian2014} proposed using IPCW to estimate
a class of linear regression models that target RMST. \citet{wang2018} 
described an approach that uses weighted estimating equations with IPCW to estimate a
collection of RMSTs in the context of dependent censoring, and
\citet{zhong2022} also proposed a similar strategy to estimate regression models for RMST at a collection of truncation points. 
\citet{xiang2012} combined the pseudo-outcome approach with IPCW to estimate the restricted mean of a log-survival time
under dependent censoring.

An advantage of using IPCW is its ability to directly handle informative censoring 
where the survival and censoring times are not independent but are conditionally independent given certain
patient covariate information. 
With IPCW, one does not need to specify a joint model for survival and censoring times as such modeling
can be quite complex and involve the specification of model features that are not directly related to the RMST.
Rather, one can model the censoring distribution separately from the objective function used for RMST estimation
and then plug the estimated censoring probabilities into an RMST-driven objective function. 

Our main goal in this work is to bridge IPCW approaches for estimating RMST with flexible Bayesian models for inference.
Combining these approaches can harness the advantages of Bayesian procedures for uncertainty quantification and
the robustness and simplicity of IPCW-based procedures.
However, a major obstacle when seeking to incorporate IPCW as part of a Bayesian procedure for performing inference 
about a collection of RMST values is that IPCW does not directly fit into the classical Bayesian framework. 
Typically, estimation of an RMST function using IPCW is performed by solving a weighted estimating equation where the uncertainty 
in the weights is ignored and where the estimating equation may or may not have any connection with an appropriate log-likelihood.
To address this, we propose basing inferences about an RMST function around the 
loss function-driven ``generalized Bayes'' framework described in \citet{bissiri2016}.
This generalized Bayes framework has recently seen increasing popularity and has 
been deployed in a variety of contexts, for example, in clustering (\citet{rigon2023}) and in
estimating the distribution of a categorical outcome under dataset shift (\citet{fiksel2022}). 
The main feature of generalized Bayes is that it provides a mechanism 
to update a prior to a posterior distribution even when one has not
specified a full joint probability model for all components involved and has only
specified a loss function of interest. The pseudo-Bayesian posterior distribution resulting 
from such a prior-updating procedure is often referred to as a ``Gibbs posterior'' (\citet{martin2021}). 
Using a generalized Bayes approach allows us to directly target the RMST function of interest without 
needing to choose a model that completely describes the distribution of the survival outcomes.
This allows us to avoid having to model nuisance parameters, which makes both
the model specificication and posterior computation less cumbersome. Moreover, it allows our 
inferences about the RMST function to be more robust to model misspecification.
Indeed, as noted in \citet{jiang2008}, when model misspecification is present, 
the performance of a likelihood-based Bayesian procedure can be sub-optimal for a risk function of interest 
even though the posterior may be consistent for a model that is closest, in the Kullback-Leibler sense to the true model.

The procedure we propose for finding a Gibbs posterior of an RMST function is not, however, an
exact application of the generalized Bayes updating procedure described in \citet{bissiri2016}.
This is because the natural IPCW loss function for RMST
includes weights that depend on features of the unknown censoring distribution,
and hence, the loss function of interest depends on the censoring distribution.
To handle this, we propose an updating procedure that can be seen as arising from two steps.
First, we form a censoring distribution-conditional Gibbs posterior for the RMST function by applying
the generalized Bayes updating rule to an IPCW-driven loss function for RMST.
Then, we form the final Gibbs posterior for RMST by averaging the censoring-conditional
Gibbs posterior with respect to a posterior distribution for the censoring distribution.

The two-step procedure described above can be applied to any prior that one chooses for the RMST function and 
any likelihood-prior model that one selects for the censoring distribution.
However, in this paper, we focus on a specific prior for the RMST function and two models for the censoring distribution. 
Specifically, we concentrate on the use of Bayesian additive regression trees (BART)
priors (\citet{chipman2010bart}) for the unknown RMST function, and we use a nonparametric gamma process prior for noninformative censoring
and a BART-based accelerated failure time model for the case of informative censoring. 
BART is a flexible nonparametric prior that represents the RMST function as an ensemble of decision trees and places a regularization prior on the components of each tree.
The main advantages of BART are that it can directly incorporate 
both continuous and categorical covariates and that BART can automatically adapt to 
nonlinearities and covariate interactions in the RMST function without requiring the user to specify 
any such interactions. We focus on BART because the emphasis
of many available methods for estimating RMST from individual-level baseline covariates has been on
more traditional regression models, and flexible nonparametric models 
which draw upon ensemble learning techniques 
can be a useful tool for improving the prediction of RMST in many contexts. 
Moreover, even though BART is closely connected to more ``black box'' ensemble methods, 
BART can still provide direct uncertainty quantification for the underlying RMST function of interest.
We refer to the procedure that combines our two-stage generalized Bayes procedure with 
a BART prior on the RMST function as RMST-BART. 

This paper is organized as follows. In Section \ref{sec:lossF}, we define the RMST of a survival outcome and describe the associated covariate-dependent RMST function. Then, in Section \ref{sec:lossF}, we describe a loss function that targets estimation of this RMST function,
and we outline how one can use a generalized Bayes framework to bring together a model for the censoring distribution and 
the RMST-targeted loss function in order to generate a posterior distribution for the RMST function of interest.
In addition, Section \ref{sec:lossF} describes the BART prior we use for the RMST function and details the types of
models that we use to handle either noninformative or informative censoring. 
Section \ref{sec:BART-Prior} discusses our default choices of all model hyperparameters, describes the selection of a loss function tuning parameter,
and outlines our procedure for performing posterior computation.
Section \ref{sec:Simulations} reports the results of two simulation studies that evaluate the RMST-BART method and 
related procedures under
scenarios with both noninformative and informative censoring mechanisms.
Section \ref{sec:Application} exhibits the uses of our method with an analysis of a breast cancer study that contains clinical and genomic covariates, and we then conclude with a brief discussion in Section \ref{sec:Diss}.

\section{Loss Function-targeted generalized BART for RMST}\label{sec:lossF}

\subsection{Data Structure and Restricted Mean Survival Time}\label{sec:RMST}

We assume that we have a study with $n$ cases or individuals with the random variable $T_{i}$ denoting the time-to-failure of an event of interest for individual $i$
and the random variable $C_{i}$ denoting a time of right-censoring for this time-to-event outcome. We observe the vector of follow-up times $\mathbf{U} = (U_{1}, \ldots, U_{n})$ and vector of event indicators $\bdelta = (\delta_{1}, \ldots, \delta_{n})$, where $U_{i} = \min\{ T_{i}, C_{i} \}$ and $\delta_{i} = I(T_{i}\leq C_{i})$. For each pair $(U_{i}, \delta_{i})$, we observe a corresponding vector $\mathbf{x}_{i}$ of baseline covariates.

Of primary interest are the restricted mean survival times (RMST) for transformed survival times $b(T_{i})$, where $b$ is a monotone increasing function.
For an individual with covariate vector $\mathbf{x}_{i}$, the RMST parameter $\mu_{\tau}(\mathbf{x}_{i})$ with restriction point $\tau$ for the transformed survival times $b(T_{i})$ is defined as the expectation of the minimum of $b(T_{i})$ and $\tau$ conditional on the value of the covariate vector $\mathbf{x}_{i}$. Specifically,
the $\tau$-RMST of $b(T_{i})$ is defined as
\begin{equation}
\mu_{\tau}( \mathbf{x}_{i} ) = E[ \min\{ b(T_{i}), \tau \} \mid \mathbf{x}_{i} ] = E[ b(U_{i}^{\tau}) \mid \mathbf{x}_{i}], \nonumber 
\end{equation}
where $U_{i}^{\tau}$ is the random variable $U_{i}^{\tau} = \min\{ U_{i}, b^{-1}(\tau) \}$. 
The connection between the $\tau$-RMST and the area under the survival function of $b(T_{i})$ from $0$ to $\tau$
is frequently noted (e.g., \citet{royston2013restricted}). Specifically, 
if $S(t|\mathbf{x}_{i}) = P\{ b(T_{i}) > t|\mathbf{x}_{i} \}$, then $\mu_{\tau}( \mathbf{x}_{i})$ can be expressed as
\begin{equation}
\mu_{\tau}( \mathbf{x}_{i} ) = \int_{0}^{\tau} S(t|\mathbf{x}_{i})dt. \nonumber 
\end{equation}

\subsection{A Loss Function for RMST}\label{sec:loss-RMST}

Suppose we are considering a function $f(\mathbf{x}_{i})$ which takes a covariate vector $\mathbf{x}_{i}$ as 
input and whose purpose is to target the $\tau$-RMST of the transformed survival time $b(T_{i})$. 
If we assumed the censoring-survival distribution conditional on $\mathbf{x}_{i}$, i.e., 
$G(t | \mathbf{x}_{i}) = P( C_{i} > t | \mathbf{x}_{i})$,
was known, a natural inverse probability of censoring weighted (IPCW) loss function
to evaluate the performance of $f(\cdot)$ is the following
\begin{eqnarray}
\mathcal{L}_{\eta}(f| \Lambda; \mathbf{U}^{\tau}, \bdelta ) 
= \eta \sum_{i=1}^{n} \delta_{i}\exp\{ \Lambda(U_{i}^{\tau}| \mathbf{x}_{i}) \}\big[ b(U_{i}^{\tau}) - f(\mathbf{x}_{i}) \big]^{2},
\label{eq:main_loss_function}
\end{eqnarray}
where $\mathbf{U}^{\tau} = \big( \min\{U_{1}, \tau\}, \ldots, \min\{ U_{n}, \tau\} \big)$ denotes the $n \times 1$ vector 
of $\tau$-truncated follow-up times and where $\eta > 0$ is a positive scalar.
In Equation (\ref{eq:main_loss_function}), $\Lambda(t|\mathbf{x}_{i})$ is the censoring cumulative hazard function 
conditional on the value of $\mathbf{x}_{i}$. Under the assumption
that the censoring survival function $G(t|\mathbf{x}_{i})$ is continuous in $t$ for each $\mathbf{x}_{i}$,
$G(t|\mathbf{x}_{i}) = \exp\{ -\Lambda(U_{i}^{\tau}| \mathbf{x}_{i}) \}$,
and hence, using weights $\exp\{ \Lambda(U_{i}^{\tau}| \mathbf{x}_{i}) \}$
is equivalent to using inverse censoring weights $1/G(t|\mathbf{x}_{i})$.

The justification for considering a loss function of the form (\ref{eq:main_loss_function}) is that,
under the assumption of independence of $T_{i}$ and $C_{i}$ given $\mathbf{x}_{i}$,
the expected value of the following expected discrepancy
\begin{equation}
E\Big\{  \delta_{i}\exp\{ \Lambda(U_{i}^{\tau}| \mathbf{x}_{i}) \}\big[ b(U_{i}^{\tau}) - \theta_{i} \big]^{2} \Big| \mathbf{x}_{i} \Big\} \nonumber 
\end{equation}
is minimized by setting $\theta_{i} = \mu_{\tau}( \mathbf{x}_{i} )$, for $i=1,\ldots, n$.
IPCW-based objective functions similar to that in equation (\ref{eq:main_loss_function}) have been used in other approaches to modeling the RMST in a regression context, including, for example, \citet{tian2014} and \citet{wang2018}.
More generally, many fundamental estimates in survival analysis can be viewed through the lens of IPCW-based
estimation; for example, as noted in \citet{satten2001}, the Kaplan-Meier estimator can be characterized as a weighted empirical distribution function using the observed event times with the inverse censoring probabilities as weights.

Our main aim is to use the IPCW-based loss function  (\ref{eq:main_loss_function}) to drive inferences about $f()$, and we do not
want our inference procedure to require any additional probability modeling of the survival times $T_{i}$.
A key challenge for achieving this within a Bayesian framework 
is that  (\ref{eq:main_loss_function}) is not a log-likelihood that corresponds to a probability model that we would want to specify for the survival times, and while (\ref{eq:main_loss_function}) is a reasonable objective function to target if one
is only interested in generating an estimate of $f(\cdot)$, classical Bayesian inference does not provide a direct procedure for 
using (\ref{eq:main_loss_function}) to update a prior distribution one may have for $f(\cdot)$. 

One approach for updating a prior distribution when one has a loss function of interest 
but does not have a likelihood is the procedure described in \citet{bissiri2016}. 
Given our loss function of interest (\ref{eq:main_loss_function}) and a prior over the unknown function $f(\cdot)$ (which we denote with $\pi_{B}$), applying the \citet{bissiri2016} procedure for updating $\pi_{B}(\cdot)$ yields the following ``Gibbs posterior'' for $f$
\begin{equation}\label{eq:fixedG_Gibbs_post}
p(f | \Lambda,\mathbf{U}^{\tau}, \bdelta) = \frac{\exp\big\{ -\mathcal{L}_{\eta}(f| \Lambda; \mathbf{U}^{\tau}, \bdelta ) \big\} \pi_{B}(f) }{ \int \exp\big\{ -\mathcal{L}_{\eta}(f| \Lambda; \mathbf{U}^{\tau}, \bdelta ) \big\} d\pi_{B}(f) }.
\end{equation}

While (\ref{eq:fixedG_Gibbs_post}) is a justifiable update of the prior $\pi_{B}( \cdot )$ when the cumulative hazard $\Lambda$ 
of the censoring times is known, we are chiefly interested in a ``Gibbs'' or ``Generalized'' posterior for 
$f(\cdot)$ which is conditional only on the observed data and does not
condition on any unknown quantities. To that end, we propose to perform inference on $f(\cdot)$
by targeting the following marginal Gibbs posterior of $f$
\begin{equation} \label{eq:Gibbs_post}
p(f|\mathbf{U}^{\tau}, \bdelta) = \int \Bigg[ \frac{\exp\big\{ -\mathcal{L}_{\eta}(f| \Lambda; \mathbf{U}^{\tau}, \bdelta ) \big\} \pi_{B}(f) }{ \int \exp\big\{ -\mathcal{L}_{\eta}(f| \Lambda; \mathbf{U}^{\tau}, \bdelta ) \big\} d\pi_{B}(f) } \Bigg] d\tilde{\pi}_{\mathbf{U}^{\tau}, \boldsymbol{\delta}}(\Lambda).
\end{equation}
In (\ref{eq:Gibbs_post}), $\tilde{\pi}_{\mathbf{U}^{\tau}, \boldsymbol{\delta}}(\cdot)$ denotes a posterior distribution over $\Lambda$ based on the data $(\mathbf{U}^{\tau}, \bdelta)$. In other words, the marginal posterior $p(f|\mathbf{U}^{\tau}, \bdelta)$ can be thought of as arising from averaging over the $\Lambda$-conditional Gibbs posteriors $p(f | \Lambda,\mathbf{U}^{\tau}, \bdelta)$. In practice, 
to sample from (\ref{eq:Gibbs_post}), one can simply take draws $\Lambda^{(s)}$ from $\tilde{\pi}_{\mathbf{U}^{\tau}, \boldsymbol{\delta}}(\cdot)$, 
and for each $\Lambda^{(s)}(\cdot)$ draw $f^{(s)}$ from the Gibbs posterior in (\ref{eq:fixedG_Gibbs_post}) using $\Lambda^{(s)}$ as the cumulative hazard function.

It is important to note that the modeling of the censoring distribution can be performed separately from the modeling of the function $f(\cdot)$. 
In that sense, our approach resembles conventional IPCW-based methods or propensity score methods in 
causal inference where the propensity score model is estimated separately from the outcome model. However, a key difference 
between our approach and conventional IPCW methods is that our weights are allowed to vary across MCMC iterations, and we do not just plug
in a single set of weights into a loss function for $f(\cdot)$.
Having variability in the weights allows for propagation of the uncertainty associated with the censoring-based weights, and
separating the modeling of censoring times from the RMST loss function allows us to avoid the ``model
feedback'' issues that could arise when trying to perform joint modeling of the RMST function and censoring
probabilities (\citet{zigler2013}). Using $\tilde{\pi}_{\mathbf{U}^{\tau}, \boldsymbol{\delta}}(\cdot)$ to generate 
multiple samples of the IPC weights instead of using a fixed set of IPC weights 
can have a notable impact on the coverage performance of the credible intervals for $f(\cdot)$. 
Appendix B shows the results of a simulation study
comparing the coverage of our approach which draws multiple samples of the IPCW
with an approach which just plugs in a single set of IPCW.

\subsection{Prior for $f$: Bayesian Additive Regression Trees}\label{sec:BART}
Bayesian additive regression trees (BART) is a flexible method originally developed to provide inferences about an unknown mean function $f(\cdot)$ that relates a vector of outcomes $\mathbf{y} \in \mathbb{R}^{n}$ with an associated design matrix of covariates $\mathbf{X} \in \mathbb{R}^{n \times p}$. 
BART generates a flexible, nonparametric prior for the mean function $f(\cdot)$ by assuming that $f(\cdot)$ 
is represented by a sum of a large collection of decision trees. Specifically, $f(\cdot)$ is assumed to have the form 
\begin{equation}\label{eq:sum-trees2}
f(\mathbf{x}) = \sum_{h=1}^{H} g_{h}(\mathbf{x}; \mathcal{T}_{h}, \bmu_{h}), 
\end{equation} 
where $H$ is the total number of pre-specified trees and $g_{h}(\mathbf{x}; \mathcal{T}_{h}, \bmu_{h})$
is the fitted value returned by the $h^{th}$ binary decision tree given an input covariate vector $\mathbf{x}$. 
The term $\mathcal{T}_{h}$ is a parameter representing the tree structure of the $h^{th}$ decision tree,
and it contains all the necessary parameters to completely define the structure of the $h^{th}$ decision tree, including 
the splitting variables and splitting values used at each interior node.
The term $\bmu_{h}$ is a vector that contains the fitted values for each terminal node of $\mathcal{T}_{h}$. 
Note that the number of elements in $\bmu_{h}$ must match the number of terminal nodes in $\mathcal{T}_{h}$.
The $h^{th}$ decision tree is fully characterized by the tree structure parameter $\mathcal{T}_{h}$
and the terminal node parameters $\bmu_{h}$.

Building on (\ref{eq:sum-trees2}), a prior distribution over $f(\cdot)$ is induced by placing
a prior distribution over the collection of individual trees: 
$\{ (\mathcal{T}_1,\bmu_1), ..., (\mathcal{T}_H,\bmu_H) \}$. 
The standard BART prior for $f(\cdot)$ assumes that $(\mathcal{T}_1,\bmu_1), ..., (\mathcal{T}_H, \bmu_H)$
are mutually independent so that the prior distribution for the collection of trees can be expressed as 
\begin{eqnarray}\label{eq:BART-Prior}
    p\big\{(\mathcal{T}_1,\bmu_1), ..., (\mathcal{T}_H,\bmu_H) \big\}
    &=&  \Bigg[ \prod_{h=1}^{H}p(\bmu_h|\mathcal{T}_h)p(\mathcal{T}_h)\Bigg ]
    = \Bigg[ \prod_{h=1}^{H}\Bigg[ \prod_{j=1}^{n_{h} }p(\mu_{hj}|\mathcal{T}_h)\Bigg ]p(\mathcal{T}_h)\Bigg ], 
\end{eqnarray}
where the vector $\bmu_h = (\mu_{h1}, ..., \mu_{hn_h})$ represents the $n_{h}$ terminal node parameters associated with tree $\mathcal{T}_h$. From equation (\ref{eq:BART-Prior}), one only needs to choose the priors for $\mu_{hj}|\mathcal{T}_{h}$ and $\mathcal{T}_{h}$ in order to fully specify the BART prior for $f(\cdot)$.

\bigskip

\noindent
{\textbf{Prior for $\bmu_{h}|\mathcal{T}_{h}$.}} For the choice of prior density $p(\mu_{hj}|\mathcal{T}_{h})$, we
use a Gaussian distribution for $\mu_{hj}$, i.e., $\mu_{hj}|\mathcal{T}_{h} \sim \textrm{Normal}(\mu_{\mu}, \sigma_{\mu}^{2})$,
as is assumed in most other implementations of BART. While this choice of prior for $\mu_{hj}$ does indeed 
place prior mass on implausible values of $\mu_{hj}$ because the $\tau$-RMST function $\mu_{\tau}(\mathbf{x})$ is restricted to lie
in between $b(0)$ and $\tau$, the prior probability assigned to implausible
values of $\mu_{hj}$ is typically very small for choices of the prior standard deviation parameter $\sigma_{u}$ that we have found to be suitable in practice. 
While a prior distribution for $\mu_{hj}$ whose support is entirely contained within $( b(0), \tau )$ 
may seem to be more sensible in this context, using a Gaussian prior has a number of computational 
advantages as the conditional distribution of $\mathcal{T}_{h}$ has a closed-form expression that does not depend on $\bmu_{h}$.
Default choices for the hyperparameters $(\mu_{\mu}, \sigma_{\mu})$ of $p(\mu_{hj}|\mathcal{T}_{h})$
are discussed in Section \ref{sec:Hyp-par}.

\bigskip

\noindent
{\textbf{Prior of $\mathcal{T}_{h}$.}} 
To fully specify the BART prior for $\mathcal{T}_{h}$, one needs to specify probability models
for the node generation process, the splitting variable selected at each internal node, and 
the splitting variable selected at each internal node. For these components, we use the standard BART prior for 
$\mathcal{T}_{h}$ outlined in \citet{chipman2010bart} with a quantile-based rule for the prior distributions of the splitting values. 
Namely, for the node generation process, 
one views the realization of the binary tree nodes as being generated 
from repeatedly splitting or terminating nodes until all nodes have been deemed terminal. 
At a given node depth $d$, a node is split upon with probability
$\alpha_{*}(1 + d)^{-\beta_{*}}$ and is terminated with probability $1 - \alpha_{*}(1 + d)^{-\beta_{*}}$.
For a tree with a given collection of terminal and interior nodes, the splitting variables used at the interior nodes
are selected from a discrete uniform distribution over the set of available splitting variables.
Similarly, given the splitting variables chosen, the splitting value for a particular splitting variable is 
selected from a discrete uniform distribution
over a grid of values placed at a pre-specified number of quantiles of the variable used.

One may also note that the standard BART regression model incorporates a residual variance parameter $\sigma^{2}$ and a corresponding prior, whereas our approach does not have such an explicit residual variance term. However, the tuning parameter $\eta$ in loss function (\ref{eq:main_loss_function})
does play a similar role to a residual variance parameter in the sense that
$\eta$ determines the relative weight given to the loss function versus the log-prior of $f(\cdot)$.
Rules for selecting the tuning parameter $\eta$ are discussed in Section \ref{sec:Hyp-par}.

\subsection{Modeling the Censoring Distribution}\label{sec:censoring-dist}

\textbf{Modeling the censoring distribution under noninformative censoring.}
If one assumes censoring is noninformative, then the cumulative hazard function for censoring does not depend on any covariate information, i.e., $\Lambda(t) = \Lambda(t|\mathbf{x}_{i})$. The approach we use for modeling $\Lambda(t)$ under the assumption of noninformative censoring follows the methodology described in \citet{sinha1997}. With this approach, one groups the data into a series of bins and places a gamma process prior (\citet{kalbfleisch1978,burridge1981}) on the increments of $\Lambda(t)$ across each bin interval. More specifically, 
if one constructs $J$ intervals of the form $(0, s_{1}], (s_{1}, s_{2}], \ldots, (s_{j-1}, s_{J}]$ using a sequence of pre-specified values $0 < s_{1} < s_{2} < \ldots < s_{J}$ and considers the increment $\lambda_{j}$ of $\Lambda(t)$ across the $j^{th}$ interval $\lambda_{j} = \Lambda( s_{j} ) - \Lambda(s_{j-1})$, then a gamma process prior on $\Lambda(t)$ with mean process $\alpha_{0}(t)$ and rate parameter $\kappa_{0}$ implies that
\begin{equation}
\lambda_{j} \sim \textrm{Gamma}\big\{ \kappa_{0}(\alpha_{0,j} - \alpha_{0,j-1}), \kappa_{0} \big\}, \nonumber 
\end{equation}
where $\lambda_{1}, \ldots, \lambda_{J}$ are independent and $\alpha_{0,j} - \alpha_{0,j-1} = \alpha_{0}(s_{j}) - \alpha_{0}(s_{j-1})$. Note that, in our parameterization of the gamma process, this implies that the prior mean and variance of $\lambda_{j}$ are $\alpha_{0,j} - \alpha_{0,j-1}$ and $1/\kappa_{0}$ respectively. 

If we let $E_{j} = \sum_{i=1}^{n} (1 - \delta_{i})I(s_{j-1} < U_{i}^{\tau} \leq s_{j})$ be the number of censoring events within
interval $(s_{j-1}, s_{j}]$ and $R_{j} = \sum_{i=1}^{n} I(U_{i}^{\tau} > s_{j-1})$ be the number at risk at $s_{j-1}$, then, as shown in 
\citet{sinha1997}, the joint posterior of the increment parameters $\lambda_{j}$ given the grouped censoring data $\mathbf{E} = (E_{1}, \ldots, E_{J})$ and $\mathbf{R} = (R_{1}, \ldots, R_{J})$ is
\begin{equation}
p(\lambda_{1}, \ldots, \lambda_{J}|\mathbf{E}, \mathbf{R})
\propto \prod_{j=1}^{J} \lambda_{j}^{\kappa_{0}(\alpha_{0,j} - \alpha_{0,j-1}) - 1}\exp\{ -\lambda_{j}(R_{j} - E_{j} + \kappa_{0}) \}\{ 1 - \exp(-\lambda_{j}) \}^{E_{j}}.
\label{eq:cens_post}
\end{equation}
In order to draw a sample from the posterior of $\Lambda$ evaluated at $U_{i}^{\tau}$ (which we denote with $\Lambda^{(t)}(U_{i}^{\tau})$), we can draw $\lambda_{1}, \ldots, \lambda_{J}$ from the density in equation (\ref{eq:cens_post}),
and set $\Lambda^{(t)}(U_{i}^{\tau}) = \frac{\lambda_{k_{(i)}}(U_{i}^{\tau} - s_{j-1})}{ s_{j} - s_{j-1}} + \sum_{k=1}^{k(i) - 1}\lambda_{k}$, where $k(i) = \{j \in \{1, \ldots, J\}: s_{j-1} < U_{i}^{\tau} \leq s_{j} \}$.

\bigskip

\noindent
\textbf{Modeling the censoring distribution under informative censoring using BART.}
When the cumulative hazard for censoring is allowed to depend on the covariate vector $\mathbf{x}_{i}$,
any Bayesian model which induces a posterior distribution over the function $\Lambda(t|\mathbf{x})$ for $0 \leq t \leq b^{-1}(\tau)$
can be incorporated into our framework for targeting the marginal Gibbs posterior in (\ref{eq:Gibbs_post}).

In our implementation, we have modeled $\Lambda(t|\mathbf{x})$ using the nonparametric AFT model
with a BART prior for the regression function described in \citet{henderson2020}. This modeling approach assumes the log-censoring times
are given by
\begin{equation}
\log C_{i} = m^{C}( \mathbf{x}_{i} ) + \xi_{i}, \nonumber 
\end{equation}
where a centered Dirichlet process mixture model prior is placed on the distribution of the 
mean-zero residual term $\xi_{i}$ and a BART prior is placed on the function $m^{C}(\cdot)$.
The above AFT model implies that the cumulative hazard functions take the form
\begin{equation}
\Lambda(t|\mathbf{x}_{i}) = -\log \Big\{ S_{\xi}\Big( \log t - m^{C}(\mathbf{x}_{i}) \Big) \Big\}, \label{eq:cumhaz_form}
\end{equation}
where $S_{\xi}( t ) = P( \xi_{i} > t)$ denotes the survival function of the residual term. 
Hence to draw one sample of $\Lambda$ from its posterior, one can
use a single posterior draw of $S_{\xi}(\cdot)$ and $m^{C}(\cdot)$ and then directly apply (\ref{eq:cumhaz_form}). 

\section{Prior Specification and Posterior Computation}\label{sec:BART-Prior}

\subsection{Choice of Hyperparameters}\label{sec:Hyp-par}

In the posterior computation, we use ``centered'' values of $b(U_{i}^{\tau})$,
and our default hyperparameter settings are tailored to the 
centered versions of the outcomes. Because the $U_{i}^{\tau}$ are outcomes
subject to right censoring, we do not perform a naive centering that simply subtracts the sample
mean of $b(U_{i}^{\tau})$. Rather, we center the $b(U_{i}^{\tau})$ by subtracting
an estimate of the RMST of $b(U_{i}^{\tau})$, unconditional on $\mathbf{x}_{i}$.
Specifically, we let $Y_{i}^{\tau}$ denote the centered version of $b(U_{i}^{\tau})$, which is defined as
\begin{equation}
Y_{i}^{\tau} = b(U_{i}^{\tau}) - \hat{\mu}_{b}, \nonumber 
\end{equation}
where $\hat{\mu}_{b}$ is the RMST estimate $\hat{\mu}_{b} = \frac{1}{n}\sum_{k=1}^{n} \delta_{k}b(U_{k}^{\tau}) / \hat{G}^{KM}(U_{k}^{\tau})$.
Here, $G^{KM}$ denotes the Kaplan-Meier estimate of the censoring distribution.

In this section, we will use $f^{c}(\mathbf{x})$ to denote the centered version of $f(\mathbf{x})$, i.e., $f^{c}(\mathbf{x}) = f(\mathbf{x}) - \hat{\mu}_{b}$. In practice, we place a BART prior on $f^{c}(\cdot)$ and construct our inferences about $f(\cdot)$
by simply adding $\hat{\mu}_{b}$ to draws from the posterior of $(f^{c}(\mathbf{x}_{1}), \ldots, f^{c}(\mathbf{x}_{n}) )^{T}$.

\bigskip

\noindent
\textbf{Default choice of $\eta$.} The term $\eta$ in the main loss function in equation (\ref{eq:main_loss_function}) 
can be thought of as a tuning parameter that controls the weight given to 
loss function (\ref{eq:main_loss_function}) relative to the log-prior distribution.
Several procedures for selecting this type of tuning parameter have been proposed recently
including \citet{syring2019}, \citet{holmes2017}, and \citet{bissiri2016}.

As a default choice of $\eta$, we begin with the ``unit information loss'' approach described in \cite{bissiri2016}. 
% wich is also based on ideas developed earlier in the objective Bayes literature (Kass and Wasserman?
This approach balances a loss function of interest with a prior-based loss by choosing $\eta$
so that, under the assumed prior distribution, the expected value of the loss function of interest 
matches the expected value of a particular log-prior loss. In our context, applying 
the \citet{bissiri2016} selection rule for a fixed cumulative hazard function $\Lambda$
yields the following choice of $\eta$
\begin{equation}
\eta = \frac{ E\big[ \log\{ \pi(\tilde{\btheta}_{x})/\pi(\btheta_{x})\} \big] }{ n E\big[ \delta_{i} \exp\{\Lambda(U_{i}^{\tau})\} \{Y_{i}^{\tau} - f^{c}(\mathbf{x}_{i})\}^{2} \big] }
= \frac{ E\big[ \log\{ \pi(\tilde{\btheta}_{x})/\pi(\btheta_{x})\} \big] }{ n E\big[ \{ b(T_{i}^{\tau}) - \hat{\mu}_{b} - f^{c}(\mathbf{x}_{i})\}^{2} \big] },
\label{eq:eta_form}
\end{equation}
where $T_{i}^{\tau} = \min\{ T_{i}, \tau \}$ and where
$\btheta_{x}$ is the $n \times 1$ vector defined as $\btheta_{x} = \big( f^{c}(\mathbf{x}_{1}), \ldots, f^{c}(\mathbf{x}_{n}) \big)^{T}$ and $\tilde{\btheta}_{x}$ is the maximizer of $\pi(\btheta_{x})$. In (\ref{eq:eta_form}), the expectation in the numerator is taken with respect to the prior distribution of $\btheta_{x}$, and the expectation in the denominator is taken with respect to the prior joint distribution of $(T_{i}^{\tau}, f^{c}(\mathbf{x}_{i}))$. 
Note that $\eta$ in (\ref{eq:eta_form}) does not depend on $\Lambda$ due to the assumed conditional independence between 
$T_{i}$ and $C_{i}$ conditional on $\mathbf{x}_{i}$. Note also that because $\hat{\mu}_{b}$ just serves to center the $b(U_{i}^{\tau})$, we regard it as a fixed constant when evaluating the expectation in the denominator of (\ref{eq:eta_form}).

To approximate the numerator of (\ref{eq:eta_form}), we use a Gaussian process approximation to the prior distribution of $\btheta_{x}$ (i.e., $\btheta_{x} \sim \textrm{Normal}(\mathbf{0}, \bSigma_{0})$ 
for some covariance matrix $\bSigma_{0}$). Under this assumption, we have that 
$\tilde{\btheta}_{x} = \mathbf{0}$, $\log\{ \pi(\tilde{\btheta}_{x})/\pi(\btheta_{x})\} = \frac{1}{2}\btheta_{x}^{T}\bSigma_{0}^{-1}\btheta_{x}$, and hence
\begin{equation}
E\Big[ \log\{ \pi(\tilde{\btheta}_{x})/\pi(\btheta_{x})\} \Big] = \frac{1}{2}E\Big( \btheta_{x}^{T}\bSigma_{0}^{-1}\btheta_{x} \Big)
= \frac{1}{2}\textrm{tr}\Big( \bSigma_{0}^{-1}\bSigma_{0} \Big) = \frac{n}{2}.
\label{eq:logprior_approx}
\end{equation}
Combining (\ref{eq:eta_form}) and (\ref{eq:logprior_approx}), yields the unit 
information loss choice of $\eta = \sigma_{r}^{2}/2$, where $\sigma_{r}^{2} = E\big[ \{ b(T_{i}^{\tau}) - \hat{\mu}_{b} - f^{c}(\mathbf{x}_{i})\}^{2} \big].$
Because the value of the expected squared residual $\sigma_{r}^{2}$ is often difficult to specify \textit{a priori},
we propose to use, as a default, the estimated residual variance $\tilde{\sigma}_{r}^{2}$ from a linear AFT model
as the choice of $\sigma_{r}^{2}$. More specifically, when the number of covariates $p$ is less than or equal to $n/5$,
we fit an AFT model with responses $\exp\{ b(U_{i}^{\tau}) \}$ with an assumed Weibull distribution for 
the values of $\exp\{ b(U_{i}^{\tau} )\}$. When $p > n/5$, we choose $\sigma_{r}^{2}$ by using the 
estimated residual variance $\tilde{\sigma}_{r}^{2}$ from a regularized semiparametric AFT model with responses $\exp\{ b(U_{i}^{\tau} )\}$, and this residual variance estimate is found using the rank-based criterion and associated fitting procedure described in \citet{suder2022}.

\bigskip

\noindent
\textbf{Choosing $\eta$ with cross-validation.} While the default choice of $\sigma_{r}^{2}$ (and hence 
the default choice of $\eta = \sigma_{r}^{2}/2$) described above often works well in 
practice, we have found that the sensitivity to the choice of $\sigma_{r}^{2}$ is substantial enough
that improved predictive performance across a wide range of scenarios can be realized
by using a more intensive data-dependent choice of $\sigma_{r}^{2}$.
To balance the advantages of using a data-adaptive choice of $\sigma_{r}^{2}$ with concerns about computational speed,
we have found that doing 5-fold cross-validation over a small number of candidate
values of $\sigma_{r}^{2}$ works quite well in practice. 
The six candidate values considered for $\sigma_{r}^{2}$ in cross-validation are 
$\{ 0.1\tilde{\sigma}_{r}^{2},~0.25\tilde{\sigma}_{r}^{2},~0.5\tilde{\sigma}_{r}^{2},~ 0.75\tilde{\sigma}_{r}^{2},~ \tilde{\sigma}_{r}^{2}, 1.5\tilde{\sigma}_{r}^{2} \}$, where
$\tilde{\sigma}_{r}^{2}$ is the default choice of $\sigma_{r}^{2}$ obtained through the procedure described above.
The objective function minimized by our cross-validation procedure is an average of IPCW estimates
of the RMST across each of the cross-validation test sets. The weights for these IPCW estimates are
found by using the Kaplan-Meier estimates of the censoring distribution on each of the test sets.

\bigskip

\noindent
\textbf{Default choice of $\sigma_{\mu}$:} As in the original formulation of BART (\citet{chipman2010bart}), our default choice of $\sigma_{\mu}$
is largely driven by the observed spread of the outcomes. Because the prior for $f^{c}(\mathbf{x})$ should reflect
its interpretation as a prediction for the transformed outcomes $Y_{i}^{\tau}$ whose values must satisfy $Y_{i}^{\tau} \leq \tau - \hat{\mu}_{b}$, we want our default choice of $\sigma_{\mu}$ to induce a prior distribution on $f^{c}(\mathbf{x})$ such that the prior probability 
assigned to the event $\{ Y_{min}^{\tau} \leq f^{c}(\mathbf{x}) \leq b(\tau) - \hat{\mu}_{b}\}$ is approximately $95\%$, where $Y_{min}^{\tau}$ is the minimum value of $Y_{i}^{\tau}$ among the observed events, i.e., $Y_{min}^{\tau} = \min(\{Y_{i}^{\tau}: \delta_{i} = 1\} )$. Because the BART prior on $f^{c}(\mathbf{x})$ implies that the
variance of $f^{c}(\mathbf{x})$, conditional on the entire set of trees, is $H\sigma_{\mu}^{2}$, choosing 
$\sigma_{\mu} = \{ b(\tau) - \hat{\mu}_{b} - b(Y_{min}^{\tau})\}/2\kappa\sqrt{H}$ with $\kappa = 2$ ensures that we 
are assigning approximately $95\%$ prior probability for $f^{c}(\mathbf{x})$ to the interval $[Y_{min}, b(\tau) - \hat{\mu}_{b}]$. 

It is worth noting that there is a degree of ``competition'' between $\sigma_{\mu}^{2}$ and $\eta$ because smaller of values 
$\sigma_{\mu}^{2}$ lead to more weight being given to the prior of $f(\cdot)$ while larger
values of $\eta$ lead to more weight being given to the loss function.
Hence, if using either cross-validation or information-based criteria for hyperparameter  tuning, trying to select
$\sigma_{\mu}$ and $\eta$ simultaneously is not likely to lead to any performance
improvements compared to either doing hyperparameter tuning of only $\sigma_{\mu}$ or performing hyperparameter
tuning of only $\eta$. In our approach, we fix $\sigma_{\mu}$ so that the prior for $f(\cdot)$ is interpretable
and only perform tuning for the loss function tuning parameter $\eta$.

\bigskip

\noindent
\textbf{Default choices of $\alpha_{*}, \beta_{*}$, and $H$.} For the hyperparameters $\alpha_{*}$ and $\beta_{*}$ 
that control the prior for the node generation process of each tree, we follow 
the recommendations of $\alpha_{*} = 0.95$ and $\beta_{*} = 2$ suggested in \citet{chipman2010bart}.
These settings of $\alpha_{*}$ and $\beta_{*}$ ensure that most of the prior probability is assigned to smaller individual trees. 
For the number of trees $H$, we set $H = 200$ as a default though, in our simulation studies, we do compare this 
with a single-tree model where $H = 1$. Setting $H=200$ generally works quite well 
as a default and is much less computationally demanding than selecting among 
multiple values of $H$ using cross-validation.

\bigskip

\noindent
\textbf{Hyperparameters of the Censoring Distribution.}
For the noninformative censoring model of Section \ref{sec:censoring-dist}, one needs to specify the hyperparameters $\kappa_{0}$ and $\alpha_{0,j} - \alpha_{0,j-1}$, for $j = 1, \ldots, J$. As a default, we set both $\kappa_{0} = 1$ and $\alpha_{0,j} - \alpha_{0,j-1} = 1$ for all $j$,
because these defaults lead to a posterior distribution for $\Lambda(\cdot)$ whose mean typically does not substantially differ from the Nelson-Aalen estimator of the cumulative hazard.
If substantial prior information about the censoring distribution is available, one may want to modify the values of $\alpha_{0,j}$
and $\kappa$ accordingly. 
As noted in \citet{ibrahim2001}, when the grid points $s_{j}$ are chosen so that the associated increment parameters $\lambda_{j}$ are not too large,
setting $\kappa_{0} = 1$ and $\alpha_{0,j} - \alpha_{0,j-1} = 1$ ensures that
the posterior of $\lambda_{j}$ in (\ref{eq:cens_post}) has an approximate Gamma distribution with mean $(E_{j} + 1)/(R_{j} - E_{j} + 1)$
which is closely related to the increments of the Nelson-Aalen estimator of the cumulative hazard.  
An additional advantage of setting $\kappa_{0} = 1$ and $\alpha_{0,j} - \alpha_{0,j-1} = 1$ is that the posterior distribution of $\lambda_{j}$ is that of the logarithm of a Beta-distributed random variable with shape parameters $R_{j} - E_{j} + 1$ and $E_{j}$
which makes drawing from this posterior distribution straightforward.

\subsection{Posterior Computation}\label{sec:Algorithm}
Our posterior computation strategy largely follows the Bayesian backfitting approach described in \citet{chipman2010bart} with the addition of an update of the cumulative hazard $\Lambda(\cdot)$ in every MCMC iteration (Algorithm \ref{alg:RMST-BART}). 
With the Bayesian backfitting approach, one sequentially updates the pair $(\mathcal{T}_{h}, \bmu_{h})$, for $h = 1, \ldots, H$, by, 
in each MCMC iteration, utilizing the distribution of $(\mathcal{T}_{h}, \bmu_{h})$ 
conditional on the observed data, the cumulative hazard $\Lambda(\cdot)$, 
and all of the other tree structures and terminal node parameters except for $(\mathcal{T}_{h}, \bmu_{h})$.
To elaborate further on this computational strategy, let us first re-express the $\Lambda$-conditional loss function (\ref{eq:main_loss_function}) for the vector of centered outcomes $\mathbf{Y}^{\tau} = (Y_{1}^{\tau}, \ldots, Y_{n}^{\tau})$ in terms of the collection of trees $\mathcal{T} = (\mathcal{T}_{1},\mathcal{T}_{2}, \ldots, \mathcal{T}_{H})$ and collection of terminal node parameters $\bmu = (\bmu_{1}, \ldots, \bmu_{H})$ for the centered regression function $f^{c}(\mathbf{x})$. Specifically, 
\begin{equation}
\mathcal{L}_{\eta}( \mathcal{T}, \bmu| \Lambda, \mathbf{Y}^{\tau}, \bdelta)
= \eta \sum_{i=1}^{n} \delta_{i}\exp\{ -\Lambda(U_{i}^{\tau}|\mathbf{x}_{i}) \}[ Y_{i}^{\tau} - f^{c}(\mathbf{x}_{i})]^{2}, \nonumber 
\end{equation}
and recall that, for a fixed value of $\Lambda(\cdot)$, 
the posterior of interest is given by
\begin{equation}
p( \mathcal{T}, \bmu| \Lambda, \mathbf{Y}^{\tau}, \bdelta)
\propto \exp\Big\{ -\mathcal{L}_{\eta}( \mathcal{T}, \bmu| \Lambda, \mathbf{Y}^{\tau}, \bdelta)  \Big\} \prod_{h=1}^{H} \pi(\bmu_{h}|\mathcal{T}_{h})\pi(\mathcal{T}_{h}).
\label{eq:gibbs_prop_density}
\end{equation}
Working from (\ref{eq:gibbs_prop_density}), the distribution of $(\mathcal{T}_{h}, \bmu_{h})$ given $\Lambda$, the observed data, all other tree structures $\mathcal{T}_{-h}$, and all other terminal node parameters $\bmu_{-h}$ is given by
\begin{multline}
p( \mathcal{T}_{h}, \bmu_{h}| \Lambda, \mathcal{T}_{-h}, \bmu_{-h}, \mathbf{Y}^{\tau}, \bdelta)
\propto \\
\exp\Big\{ -\eta \sum_{i=1}^{n} \delta_{i}\exp\{ \Lambda(U_{i}^{\tau}|\mathbf{x}_{i}) \}[ g_{h}(\mathbf{x}_{i}) - R_{ih} ]^{2} \Big\}\pi(\mathcal{T}_{h}) \pi(\bmu_{h}|\mathcal{T}_{h} ), \label{eq:gibbs_joint_conditional}  
\end{multline}
where $g_{-h}(\mathbf{x}_{i}) = \sum_{k\neq h} g_{k}(\mathbf{x}_{i};\mathcal{T}_{h}, \bmu_{h})$ and $R_{ih} = Y_{i}^{\tau} - g_{-h}(\mathbf{x}_{i})$ implying that $R_{ih}$ is the $i^{th}$ residual obtained from using all trees except the $h^{th}$ tree to construct the regression function. Integrating out the terminal node parameters of (\ref{eq:gibbs_joint_conditional}) yields the following distribution of $\mathcal{T}_{h}$
conditional on $\Lambda$, $\mathcal{T}_{-h}$, and $\bmu_{-h}$
\begin{multline}
p(\mathcal{T}_{h} | \Lambda, \mathcal{T}_{-h}, \bmu_{-h}, \mathbf{Y}^{\tau}, \bdelta)
\propto \\
\pi(\mathcal{T}_{h}) \int \exp\Big\{ -\eta \sum_{i=1}^{n} \delta_{i}\exp\{ \Lambda(U_{i}^{\tau}|\mathbf{x}_{i}) \}[ g_{h}(\mathbf{x}_{i}) - R_{ih} ]^{2} \Big\} \pi(\bmu_{h} | \mathcal{T}_{h} ) d\bmu_{h}. 
\label{eq:Tmarg_dist}
\end{multline}
Note that $g_{h}(\mathbf{x}_{i}) = \sum_{j=1}^{n_{h}}I_{jh}(\mathbf{x}_{i})\mu_{hj}$, where $I_{jh}(\mathbf{x}_{i}) = 1$
if $\mathbf{x}_{i}$ is assigned to the $j^{th}$ terminal node of $\mathcal{T}_{h}$ and $I_{jh}(\mathbf{x}_{i}) = 0$ otherwise.
Hence, because the distribution $\bmu_{h}|\mathcal{T}_{h}$ is assumed to be multivariate Gaussian,
the expression in (\ref{eq:Tmarg_dist}) has a direct, closed form.

To update the $h^{th}$ tree structure in the $t^{th}$ MCMC iteration, a proposal tree $\mathcal{T}'$ is drawn
using the ``grow'', ``prune'', and ``change'' move scheme described in \citet{chipman1998bayesian},
and the proposal $\mathcal{T}'$ is accepted or rejected using the Metropolis-Hastings ratio,
which can be directly computed using (\ref{eq:Tmarg_dist}). The update of $\mathcal{T}_{h}$ is immediately followed 
by an update of the terminal node parameters of the $h^{th}$ tree $\bmu_{h}$.
To update $\bmu_{h}$ in the $t^{th}$ MCMC iteration, one uses the fact that, conditional 
on $\Lambda$, $\mathcal{T}_{h}$, $\mathcal{T}_{-h}$, and $\bmu_{-h}$, the distribution of $\bmu_{h}$ is multivariate Gaussian
where the conditional mean of $\bmu_{h}$ is a weighted mean of the residuals $R_{ih}$ that is shrunken towards zero. Specifically,  
\begin{eqnarray}\label{eq:muPrior}
\bmu_{h}|\Lambda, \mathcal{T}_{h}, \mathcal{T}_{-h}, \bmu_{-h}, \mathbf{Y}^{\tau}, \bdelta &\sim&  \textrm{Normal}\Big(\mathbf{d}_{h} , \mathbf{D}_{h} \Big), 
\label{eq:mu_post}
\end{eqnarray}
where $\mathbf{d}_{h}$ is an $n_{h} \times 1$ vector and $\mathbf{D}_{h}$ is an $n_{h} \times n_{h}$ diagonal matrix.
The $j^{th}$ diagonal element $D_{jj,h}$ of $\mathbf{D}_{h}$ and the $j^{th}$ component $d_{j,h}$ of $\mathbf{d}_{h}$ are given by
\begin{eqnarray}
D_{jj,h} &=& \frac{1}{\eta}\Big( (2\eta\sigma_{\mu}^{2})^{-1} +  \sum_{i=1}^{n} \delta_{i}\exp\{\Lambda(U_{i}^{\tau}|\mathbf{x}_{i})\}I_{jh}(\mathbf{x}_{i}) \Big)^{-1}
\nonumber \\
d_{j,h} &=& \eta D_{jj,h}\sum_{i=1}^{n} \delta_{i}\exp\{\Lambda(U_{i}^{\tau}|\mathbf{x}_{i})\}I_{jh}( \mathbf{x}_{i} )R_{ih}. \nonumber 
\end{eqnarray}
More details about the derivation of the conditional distribution of $\bmu_{h}$ can be found in Appendix A.

Let $\mathcal{T}_{h}^{(t)}$ and $\bmu_{h}^{(t)}$ denote the values of $\mathcal{T}_{h}$ and $\bmu_{h}$
in the $t^{th}$ MCMC iteration respectively.
After updating $(\mathcal{T}_{h}^{(t)}, \bmu_{h}^{(t)})$ to $(\mathcal{T}_{h}^{(t+1)}, \bmu_{h}^{(t+1)})$
for $h = 1, \ldots, H$, one draws a new cumulative hazard $\Lambda^{(t+1)}(\cdot)$ using either 
the noninformative or informative censoring models described in Section \ref{sec:censoring-dist}.
Note that the draw of $\Lambda^{(t+1)}(\cdot)$ can be performed as described in Section \ref{sec:censoring-dist}
because the posterior distribution of $\Lambda$ used in the target posterior (\ref{eq:Gibbs_post}) does not depend on 
the centered function $f^{c}(\cdot)$ in any way. The new cumulative hazard function $\Lambda^{(t+1)}(\cdot)$ will then 
be used in updates of $(\mathcal{T}_{h}^{(t+1)}, \bmu_{h}^{(t+1)})$ in the subsequent MCMC iteration.

The steps needed to perform a single update of the cumulative hazard $\Lambda(\cdot)$, the collection of tree structures $\mathcal{T}_{1}, \ldots, \mathcal{T}_{H}$, and terminal node parameters $\bmu_{1}, \ldots, \bmu_{H}$ are summarized in
Algorithm \ref{alg:RMST-BART}.

\begin{algorithm}
\caption{RMST-BART: One MCMC iteration to update $\Lambda$, $\mathcal{T}$, and $\bmu$}\label{alg:RMST-BART}
\begin{algorithmic}
\State \textbf{Input:} $\mathbf{Y}^{\tau}$, $\mathbf{X}$, $\Lambda^{(t)}(\cdot)$, $(\mathcal{T}_{h}^{(t)}, \bmu_{h}^{(t)})_{h=1}^{H}$
\For {\texttt{$h = 1, ..., H$}}
    \State \texttt{1. Set: $R_{ih} = Y_{i}^{\tau} - \sum_{k = 1}^{h-1} g_{k}(\mathbf{X}_{i}; \mathcal{T}_{h}^{(t+1)}, \bmu_{h}^{(t + 1)}) - \sum_{k = 1}^{h+1} g_{k}(\mathbf{X}_{i}; \mathcal{T}_{h}^{(t)}, \bmu_{h}^{(t)})$, for $i = 1, ..., N$.}
    \State \texttt{2. Generate a proposal $\mathcal{T}'$ using either a grow, change, or prune move from $\mathcal{T}_{h}$}.
    \State \texttt{3. Using (\ref{eq:gibbs_prop_density}), compute the Metropolis-Hastings ratio $\alpha_{MH}$ for $\mathcal{T}'$ versus $\mathcal{T}_{h}^{(t)}$.}
    \State \texttt{4. Generate $u \sim $ Unif$(0, 1)$.}
    \State \texttt{5. If $u \leq \alpha_{MH}$, set $\mathcal{T}_{h}^{(t+1)} = \mathcal{T}'$; otherwise, set $\mathcal{T}_{h}^{(t+1)} = \mathcal{T}_{h}^{(t)}$.}
    \State \texttt{6. Draw $\bmu_{h}^{(t+1)}$ from the distribution in Equation (\ref{eq:mu_post}). }
\EndFor \textbf{end for}
\State \texttt{7. Update $\Lambda^{(t+1)}(\cdot)$ by sampling from the noninformative or informative censoring model described in Section \ref{sec:censoring-dist}.}
\end{algorithmic}
\end{algorithm}

\section{Simulations}\label{sec:Simulations}

In this section, we study the performance of the RMST-BART procedure with two simulation studies. In the first simulation study, we generate survival outcomes from a distribution whose mean is determined by the Friedman function. In our second simulation study, we draw survival outcomes from a distribution whose mean is the absolute value of a linear model where the
predictors are generated from a multivariate normal distribution with a 
first-order autoregressive (AR(1)) \citet{mills1990time} correlation structure. 
We mainly evaluate the performance of RMST-BART and competing methods on their ability to estimate the RMST function $\mu_{\tau}(\mathbf{x}_{i}')$
on a ``test set'' of $\mathbf{x}_{i}'$ values. Specifically, we assess estimation performance by recording the average root-mean squared error 
for $\mu_{\tau}(\mathbf{x}_{i}')$ across simulation replications. The average RMSE performance of an estimator
$\hat{\mu}_{\tau}$ was measured with the following quantity 
\begin{equation}
\frac{1}{S}\sum_{s=1}^{S} \sqrt{\frac{1}{n_{test}}\sum_{i=1}^{n_{test}} \{ \hat{\mu}_{\tau}( \mathbf{x}_{i}^{',s} ) - \mu_{\tau}(\mathbf{x}_{i}^{',s}) \}^{2}}, \nonumber 
\end{equation}
where $S$ is the total number of simulation replications for a particular simulation scenario, 
$n_{test}$ is the number of covariate vectors in the test set, and $\mathbf{x}_{i}^{',s}$ is the draw of
the $i^{th}$ test set vector in the $s^{th}$ simulation replication. 
In addition to evaluating RMSE, we evaluated, for a subset of the simulation scenarios, 
the coverage proportions of the credible intervals for $\mu_{\tau}( \mathbf{x}_{i}' )$ generated
by our method.

\subsection{Friedman Function}\label{sec:Sim-Friedman}
 For this simulation study, we draw survival outcomes $T_{i}$ from the following distribution
\begin{equation}
T_{i} \sim \textrm{Gamma}\Big( f(\mathbf{x}_{i})\{ 1 + f(\mathbf{x}_{i})\},   \{1 + f(\mathbf{x}_{i})\} \Big), 
\label{eq:sim_gamma} 
\end{equation}
where $\{ 1 + f( \mathbf{x}_{i} ) \}$ represents a rate parameter implying that $E\{ T_{i} \mid f(\mathbf{x}_{i}) \} = f(\mathbf{x}_{i})$ 
and $\textrm{Var}\{ T_{i} \mid f(\mathbf{x}_{i}) \} = f(\mathbf{x}_{i})\{ 1 + f(\mathbf{x}_{i})\}$. The assumed distribution of survival times in equation (\ref{eq:sim_gamma}) implies that 
the $\tau$-RMST function for $T_{i}$ is given by
\begin{eqnarray}
\mu_{\tau}( \mathbf{x}_{i} ) &=& f(\mathbf{x}_{i})\Big[ F_{gam}\big( \tau; 1 + f(\mathbf{x}_{i})\{ 1 + f(\mathbf{x}_{i}) \}, \{ 1 + f(\mathbf{x}_{i}) \} \big) \Big] \nonumber \\
&+& \tau\Big[ 1 - F_{gam}\big( \tau; f(\mathbf{x}_{i})\{ 1 + f(\mathbf{x}_{i}) \}, \{ 1 + f(\mathbf{x}_{i}) \} \big) \big) \Big], \nonumber 
\end{eqnarray}
where $F_{gam}(x; \alpha, \beta)$ denotes the cumulative distribution function of a Gamma random variable with shape parameter $\alpha$ and rate parameter $\beta$. In equation (\ref{eq:sim_gamma}), we assume that $\mathbf{x}_{i} \in \mathbb{R}^{p}$ (with $p > 5$) and $f(\mathbf{x}_{i})$ is 
the Friedman function (\citet{friedman1991multivariate}) which is defined as
\begin{equation}
f(\mathbf{x}_{i}) = 10\sin(\pi x_{i1}x_{i2})+20(x_{i3}-0.5)^2 + 10x_{i4} +5x_{i5}. 
\label{eq:friedman_fn}
\end{equation}
Note that the Friedman function $f(\mathbf{x}_{i})$ only depends on the first five components of $\mathbf{x}_{i}$ and not on the last $p - 5$ components. In our simulation study, we draw the components $x_{ij}$ of $\mathbf{x}_{i} = (x_{i1}, \ldots, x_{ip})^{T}$ independently from a uniform distribution over $(0, 1)$, i.e., $x_{ij} \sim \textrm{Uniform}(0, 1)$.
For the distribution of the censoring times $C_{i}$, we considered two situations: one in which censoring is
noninformative, and one in which censoring is informative but $T_{i}$ and $C_{i}$ are independent given $\mathbf{x}_{i}$.

\bigskip 

\noindent
\textbf{Noninformative Censoring.}
In this case, the censoring times $C_{i}$ are independently sampled from a gamma distribution, i.e., $C_i \sim \textrm{Gamma}(3.2, r)$, and 
we consider two different values of $r$, $r = 0.1$ and $r = 0.2$. 
When $r = 0.1$, the censoring rate is relatively low with roughly $10 - 20\%$ of observations being censored
in every simulation replication, and when $r = 0.2$, the censoring rate is higher with a roughly $40 - 50\%$
the censoring rate in each simulation replication. For the noninformative censoring simulation scenarios, we set $\tau = 25$
as the restriction point of RMST because $\tau = 25$ ensures that $P(C_{i} > \tau)$ is relatively small when $r = 0.2$.
Specifically, when $r = 0.2$ and $\tau = 25$, $P(C_{i} > \tau)$ is approximately $0.15$.

\bigskip

\noindent
\textbf{Informative Censoring.} For these scenarios, censoring times $C_{i}$ and $T_{i}$
are independent conditional on $\mathbf{x}_{i}$, but the distributions of both $T_{i}$ and $C_{i}$ depend on $\mathbf{x}_{i}$.
In particular, the censoring times are drawn from the following distribution
\begin{equation}
C_{i} \sim \textrm{Gamma}\{r_{D}, 0.01 f(\mathbf{x}_{i}) \}, \label{eq:inf_cens_sim}
\end{equation}
where $f(\mathbf{x}_{i})$ is the Friedman function defined in (\ref{eq:friedman_fn}).
The shape parameter $r_{D}$ is varied across two levels $r_{D} = 1$ and $r_{D} = 3$.
When $r_{D} = 1$, the censoring is quite ``heavy'' with roughly $80\%$ observations
being censored, and when $r_{D} = 3$, the censoring 
is more moderate with a roughly $40\%$ censoring rate. The reason for only considering scenarios with a $\geq 40\%$ 
censoring rate is to highlight performance differences between using weights $\exp\{ \Lambda(U_{i}^{\tau}) \}$ which
do not depend on $\mathbf{x}_{i}$ and in using weights $\exp\{ \Lambda(U_{i}^{\tau}|\mathbf{x}_{i}) \}$ which
do depend on $\mathbf{x}_{i}$, and in our experience, there is very little performance difference between
these two approaches whenever censoring is very light (e.g., a $10 - 20\%$ censoring rate). For the informative censoring simulation scenarios, we used the same choice of restriction point (i.e., $\tau = 25$) as in the noninformative censoring scenarios. The justification for choosing $\tau = 25$ is that $P(C_{i} > \tau)$ is quite small when $r_{D} = 1$; specifically, $P(C_{i} > \tau) \approx 0.05$.

\subsection{Absolute Value Linear Model with Correlated Predictors}\label{sec:Sim-AR1}
In this simulation study, we again simulated survival outcomes
from the following distribution 
\begin{equation}
T_{i} \sim \textrm{Gamma}\Big( f(\mathbf{x}_{i})\{ 1 + f(\mathbf{x}_{i})\},   \{1 + f(\mathbf{x}_{i})\} \Big), \nonumber 
\end{equation}
but we chose $f(\mathbf{x}_{i})$ to be a direct function of a linear combination of elements in $\mathbf{x}_{i}$.
For the function $f(\mathbf{x}_{i})$, we assume that it equals the absolute value of 
a linear combination of the $p > 5$ covariates $\mathbf{x}_{i} = (x_{i1}, \ldots, x_{ip})^{T}$ 
with only the first $5$ of these covariates having nonzero coefficients in this linear combination. Specifically,
\begin{equation}
f(\mathbf{x}_{i}) = \Big| \sum_{j=1}^{5} \beta_{j}x_{ij} \Big|. \nonumber 
\end{equation}
The vector $\mathbf{x}_{i}$ is generated from a multivariate normal distribution with mean $0$ and where 
the components of $\mathbf{x}_{i}$ have an AR(1) correlation structure with an autoregressive parameter equal to $0.5$. 
That is, $\textrm{Corr}( x_{ij}, x_{ik} ) = 0.5^{|k-j|}$ and $\textrm{Var}(x_{ij}) = 1$.

In this simulation study, we only examined performance for the case of noninformative censoring with censoring times $C_{i}$ generated independently from $T_{i}$ using the distribution $C_{i} \sim \textrm{Gamma}(2.2, r)$. 
We varied $r$ across two levels: $r = 1.8$ and $r = 0.8$. The value $r = 1.8$ was utilized to achieve a lower censoring rate (i.e., a roughly $10-20\%$ censoring rate) while $r = 0.8$ was used to yield a higher censoring rate among the 
simulated observations (i.e., a roughly $40-50\%$ censoring rate). We set the truncation point $\tau$ to $5$ in order to yield a relatively small value of $P(C_{i} > \tau)$ when $r = 0.8$. Specifically, when $\tau = 5$ and $r = 0.8$, $P(C_{i} > \tau) \approx 0.10$.

\subsection{Simulation Results}

In both simulation studies, we generated a ``training set'' of observations $\{ (\mathbf{x}_{i}, U_{i}, \delta_{i}); i = 1, \ldots, n\}$
along with an independent ``test set'' of covariate vectors $\{ \mathbf{x}_{i}'; i=1, \ldots, n_{test}\}$, and for each simulation study,
we considered two different values for the number of observations in the training set: $n=250$ and $n=1000$. In each simulation replication,
we evaluate performance with $n_{test} = 1000$ separate $\mathbf{x}_{i}'$ values in the test set. 
In the Friedman function simulation study with noninformative censoring, we explored two settings of the number of covariates: $p = 10$ and $p = 100$. In both the informative censoring scenarios of the Friedman function simulation study and in the linear model simulation study, we considered $p=10$ and $p=50$.

In both simulation studies, the average root mean squared error (RMSE) of our RMST-BART approach is compared (Tables \ref{tab:friedman}, \ref{tab:friedmanDependent}, and \ref{tab:linearAR1}) with the following related methods: RMST-BART with a single tree, which we refer to as RMST-BCART due to its connection with Bayesian classification and regression trees (\citet{chipman1998bayesian}); a Cox proportional hazards regression model (CoxPH) (\citet{cox1972regression}); an $L_{1}$-penalized Cox proportional hazards regression model (Penalized CoxPH) (\citet{tibshirani1997lasso}), inverse probability of censoring weighted boosting (IPCW-boost) (\citet{hothorn2006survival}); a log-normal AFT model (\citet{wei1992accelerated}) with only an intercept term (AFT-Null); a log-normal AFT model where the regression function for log-survival times is a linear combination of all covariates (AFT-Linear); and an AFT-based BART model (AFT-BART) (\citet{chipman2010bart} and \citet{sparapani2021nonparametric}).
The AFT-BART model assumes an AFT model for the survival times while placing a BART prior on the regression function and using a Gaussian distribution for the residual of the log-survival times. Note that our simulation results present two versions of the RMST-BART, RMST-BCART, and AFT-BART approaches;
one which includes the ``-default'' suffix and one which has no suffix. The ``-default'' versions of RMST-BART and RMST-BCART
use the default choices of $\eta$ while the versions of RMST-BART and RMST-BCART without the ``-default'' suffix 
use the value of $\eta$ found through five-fold cross-validation. The default version of AFT-BART uses all hyperparameter
defaults employed in the \verb"BART" package in \verb"R" (\citet{sparapani2021nonparametric}),
and AFT-BART performs five-fold cross-validation, as suggested in \citet{chipman2010bart}, with respect to 
three different candidate pairs of values for the hyperparameters of the residual variance parameter.

Table \ref{tab:friedman} summarizes the results from the Friedman function simulation study for the noninformative censoring case.
The results presented in Table \ref{tab:friedman} indicate that, in all scenarios, our proposed method, RMST-BART, yields a lower average 
RMSE than the competing methods. In scenarios with light censoring, i.e., those with $r = 0.1$,
AFT-BART is consistently quite competitive with RMST-BART, but in scenarios where censoring is heavier, the performance gap
between AFT-BART and RMST-BART is more substantial.
When $n = 250$, the two variations of RMST-BART, i.e., 
RMST-BART with the default choice of $\eta$ and RMST-BART with a cross-validation choice of $\eta$, perform quite similarly, 
but when $n = 1000$, the larger sample size provides a clear advantage to using a cross-validation choice of $\eta$.

\begin{table}
\small
\centering
\begin{tabular}[h]{lllllllll}
\toprule
\multicolumn{1}{c}{\# of observations} &  \multicolumn{4}{c}{$n=250$} & \multicolumn{4}{c}{$n=1000$}\\
\cmidrule(l{3pt}r{3pt}){2-5}\cmidrule(l{3pt}r{3pt}){6-9}
\multicolumn{1}{c}{\# of predictors} & \multicolumn{2}{c}{$p=10$} & \multicolumn{2}{c}{$p=100$}& \multicolumn{2}{c}{$p=10$} & \multicolumn{2}{c}{$p=100$}\\
\cmidrule(l{3pt}r{3pt}){2-3}\cmidrule(l{3pt}r{3pt}){4-5}\cmidrule(l{3pt}r{3pt}){6-7}\cmidrule(l{3pt}r{3pt}){8-9}
Method& $r=0.1$& $r=0.2$ & $r=0.1$& $r=0.2$ & $r=0.1$& $r=0.2$ & $r=0.1$& $r=0.2$ \\

\midrule
 RMST-BART-default &1.48&2.14&2.46&3.15&1.10&1.41&1.33&1.66\\
 RMST-BART &1.54&2.10&2.37&2.82&0.62& 0.83&1.08& 1.64\\
RMST-BCART-default  &3.95&4.29&5.19&5.88&3.70&4.07&4.96&4.85\\
RMST-BCART  &3.86&4.46&5.11&6.01&3.59&4.10&4.78&4.97\\
 CoxPH &3.30& 2.93& 4.05&4.43&2.83&2.62&2.90&2.84\\
 Penalized CoxPH & 3.39& 3.00&3.67&3.22&2.86&2.65&2.88&2.72\\
 IPCW-boost  &2.54&2.84&2.61&2.96&2.50&2.78&2.50&2.80\\
 AFT Null  &5.38&5.34&5.35&5.35& 5.34&5.35&5.34&5.33\\
 AFT-Linear &2.81&3.05&3.80& 4.84&2.76& 3.03&2.87&3.12\\
 AFT-BART-default  &1.71&3.16&2.61& 4.32&1.12&2.85&1.57&3.26\\
  AFT-BART  &1.70& 3.17&2.61&4.31& 1.12&2.85&1.57&3.26\\
 \bottomrule
\end{tabular}
\caption{Friedman function simulation study. Average RMSE results for RMST-BART and related approaches. Censoring is noninformative with censoring times generated from the $\textrm{Gamma}(3.2, r)$ distribution. Two values are used for $r$: $0.1$ and $0.2$, which result in approximately $10-20\%$ and $40-50\%$ censoring rates, respectively.}
\label{tab:friedman} 
\end{table}

Table \ref{tab:friedmanCOV} shows test-set coverage results in the Friedman function study for $95\%$ credible intervals from the following four methods: RMST-BART-default, RMST-BART (i.e., selecting $\eta$ using cross-validation), RMST-BCART, and AFT-BART.
Here, average coverage of the RMST function $\mu_{\tau}(\mathbf{x}_{i})$ is computed by looking at the average coverage 
across points $\mathbf{x}_{i}'$ in the test set. Specifically, we measure average coverage with
\begin{equation}
\frac{1}{S}\sum_{s=1}^{S} \frac{1}{n_{test}}\sum_{i=1}^{n_{test}} I\Big\{ \hat{\mu}_{\tau}^{L}( \mathbf{x}_{i}^{',s} ) \leq \mu_{\tau}(\mathbf{x}_{i}^{',s}) \leq \hat{\mu}_{\tau}^{L}( \mathbf{x}_{i}^{',s} ) \Big\}, \nonumber 
\end{equation}
where $\hat{\mu}_{\tau}^{L}( \mathbf{x}_{i}^{',s} )$ and $\hat{\mu}_{\tau}^{U}( \mathbf{x}_{i}^{',s} )$ denote the lower and upper bounds of a $95\%$ credible interval for $\mu_{\tau}(\mathbf{x}_{i}^{',s})$ respectively and where $\mathbf{x}_{i}^{',s}$ is the draw of
the $i^{th}$ test set vector in the $s^{th}$ simulation replication.
For most scenarios presented in Table \ref{tab:friedmanCOV}, the RMST-BART method has coverage which is fairly close to $0.95$ with coverage closer to $0.95$ than AFT-BART in all but one of the scenarios. The only case where RMST-BART coverage is notably less than $0.95$ is when censoring
is heavy and the number of parameters is relatively large compared to the sample size, i.e., when $p = 100$, $r =0.2$, and $n=250$, 
but the coverage for $p = 100$ and $r = 0.2$ improves notable when the sample size is increased to $n = 1000$.
Expectedly, the RMST-BART procedure with a single tree (i.e., RMST-BCART) typically has poor coverage
as the single-tree model prior is heavily concentrated on relatively simple forms for the RMST function. 
While the choice of hyperparameters and the selection of $\eta$ are not explicitly constructed
to target nominal frequentist coverage, our experience is that the RMST-BART $95\%$ credible intervals
generally provide good coverage whenever both censoring is not extremely high and when the number of covariates
is not very large relative to sample size. 

\begin{table}
\small
\centering
\begin{tabular}[h]{lllllllll}
\toprule
\multicolumn{1}{c}{\# of observations} &  \multicolumn{4}{c}{$n=250$} & \multicolumn{4}{c}{$n=1000$}\\
\cmidrule(l{3pt}r{3pt}){2-5}\cmidrule(l{3pt}r{3pt}){6-9}
\multicolumn{1}{c}{\# of predictors} & \multicolumn{2}{c}{$p=10$} & \multicolumn{2}{c}{$p=100$}& \multicolumn{2}{c}{$p=10$} & \multicolumn{2}{c}{$p=100$}\\
\cmidrule(l{3pt}r{3pt}){2-3}\cmidrule(l{3pt}r{3pt}){4-5}\cmidrule(l{3pt}r{3pt}){6-7}\cmidrule(l{3pt}r{3pt}){8-9}
Method& $r=0.1$& $r=0.2$ & $r=0.1$& $r=0.2$ & $r=0.1$& $r=0.2$ & $r=0.1$& $r=0.2$ \\
\midrule
 AFT-BART-default &0.91&0.91&0.90&0.79&0.93&0.88&0.92&0.86\\
 RMST-BART &0.97&0.97&0.98&0.77&0.94&0.93&0.93&0.88\\
RMST-BCART  &0.67&0.49&0.59&0.57&0.77&0.70&0.58& 0.59\\
 AFT-BART &0.91&0.83&0.90&0.79&0.93&0.88&0.92&0.86\\
 \bottomrule
\end{tabular}
\caption{Average test set coverage of $95\%$ credible intervals in the Friedman function simulation study with noninformative censoring. Coverage of credible intervals from the following methods are shown: RMST-BART-default, RMST-BART, RMST-BCART, and AFT-BART.}
\label{tab:friedmanCOV} 
\end{table}

Table \ref{tab:friedmanDependent} compares the average RMSEs for the Friedman function simulation study when 
censoring times are drawn from the informative censoring model (\ref{eq:inf_cens_sim}). 
This table contains two settings of the censoring distribution; one with a ``medium'' censoring rate where $r_{D} = 3$ 
and another with a ``heavy'' censoring rate where $r_{D} = 1$. In the scenarios shown in Table \ref{tab:friedmanDependent}, RMST-BART with covariate-dependent modeling of the IPCW (i.e., RMST-BART-DEP) usually
has somewhat better RMSE performance than RMST-BART, particularly in the heavy censoring scenarios where $r_{D} = 1$. 
The performance of RMST-BART-DEP-default which uses the default choice of $\eta$ does seem to be more sensitive to the 
particular simulation scenario with rather lackluster performance when $r_{D} = 3$. 
This table also shows the impact of smaller sample sizes on the performance of RMST-BART when the covariate-dependent censoring model is used.
When $n=250$ both RMST-BART and RMST-BART-DEP are outperformed by several other methods including, for example, regularized CoxPH
and IPCW boosting. It is likely that when $n = 250$ and, especially, when censoring is heavy 
the posterior draws of the IPC weights from the covariate-dependent censoring model are quite noisy,
and it is only when $n=1000$ that the model learns these weights effectively enough
to deliver stronger predictive performance.

\begin{table}
\small
\centering
\begin{tabular}[h]{lllllllll}
\toprule
\multicolumn{1}{c}{Correlation} &  \multicolumn{4}{c}{$n = 250$} & \multicolumn{4}{c}{$n = 1000$}\\
\cmidrule(l{3pt}r{3pt}){2-5}\cmidrule(l{3pt}r{3pt}){6-9}
\multicolumn{1}{c}{\# of predictors} & \multicolumn{2}{c}{$p=10$} & \multicolumn{2}{c}{$p=50$}& \multicolumn{2}{c}{$p=10$} & \multicolumn{2}{c}{$p=50$}\\
\cmidrule(l{3pt}r{3pt}){2-3}\cmidrule(l{3pt}r{3pt}){4-5}\cmidrule(l{3pt}r{3pt}){6-7}\cmidrule(l{3pt}r{3pt}){8-9} 
Method& $r_{D}=3$& $r_{D} = 1$ & $r_{D} = 3$ & $r_{D}=1$  & $r_{D}=3$& $r_{D}=1$  & $r_{D}=3$& $r_{D}=1$ \\
\midrule 
 RMST-BART-DEP-default&4.94&3.03&5.25&5.60&3.14&1.80&3.52&2.70 \\
 RMST-BART-DEP&3.46&3.15&3.84&5.73&1.79&1.89&2.26&2.92 \\
 RMST-BART-default&4.29&3.60&4.82&6.20&2.69&1.94&3.07&3.20 \\
 RMST-BART&3.34&3.63&4.49&6.08&1.81&2.08&2.55&3.33 \\
 Coxph&3.07&3.02&3.39&5.14&2.74&2.62&2.83&2.88 \\
 Regularized Coxph&3.18&2.82&3.34&3.21&2.78&2.61&2.87& 2.70 \\
 IPCW-boost&2.70&3.34&2.79&3.90&2.63&3.16&2.65&3.30 \\
 AFT Null&5.05&4.91&5.06&4.89&5.03&4.90&5.06&4.89 \\
 AFT-Linear & 3.02&6.37&3.39&7.35&2.86&6.26&3.13&6.57 \\
 AFT-BART-default & 2.54&6.83&3.28&6.97&2.26&7.40&2.45&7.61 \\
 AFT-BART & 2.54&6.83&3.28&6.97&2.26&7.40&2.45&7.60 \\
 \bottomrule
\end{tabular}
\caption{Friedman function simulation study with informative censoring. Average RMSE results for RMST-BART and related approaches. Censoring times are generated using model (\ref{eq:inf_cens_sim}).  ``Medium'' and ``heavy'' censoring rates correspond to the $r_{D} = 3$ and $r_{D} = 1$ settings, respectively.} 
\label{tab:friedmanDependent} 
\end{table}

Table \ref{tab:friedmanKappa} shows a brief sensitivity analysis examining the impact of the choice
of the key hyperparameter $\kappa$. As described in Section \ref{sec:Hyp-par}, the standard deviation $\sigma_{\mu}$ of 
the terminal node parameters is proportional to $1/\kappa$, and our default choice of $\kappa$ is
$\kappa = 2$. Our sensitivity analysis compares RMSE performance for four values of $\kappa$: $0.25$, $0.5$, $2$, and $4$. 
As shown in Table \ref{tab:friedmanKappa}, using cross-validation to select $\eta$ greatly reduces the sensitivity to
the choice of $\kappa$ as $\eta$ can adaptively upweight or downweight the loss function if the prior for the terminal 
node parameters leads to too much or too little shrinkage. However, using the default choice of $\eta$ as is done in the RMST-BART-default 
method makes the specific choice of $\kappa$ more meaningful. As shown in Table \ref{tab:friedmanKappa}, varying $\kappa$
from $\kappa = 0.5$ up to $\kappa = 4$ can have a considerable impact on RMSE performance though changing $\kappa$ from 
$\kappa = 0.5$ to $\kappa = 2$ has a notable, but not as considerable of an impact. Though certainly not guaranteed to 
always be the best choice when using RMST-BART-default, the default choice of $\kappa = 2$ has the best performance among the scenarios
used in Table \ref{tab:friedmanKappa}, and we have found, in our experience, 
that setting $\kappa = 2$ serves as a good default choice across a range of other simulation experiments.

\begin{table}
\small
\centering
\begin{tabular}[h]{llllllllll}
\toprule
\multicolumn{1}{c}{\# of observations} &\multicolumn{1}{c}{}&  \multicolumn{4}{c}{$n=250$} & \multicolumn{4}{c}{$n=1000$}\\
\cmidrule(l{3pt}r{3pt}){3-6}\cmidrule(l{3pt}r{3pt}){7-10}
\multicolumn{1}{c}{\# of predictors} & \multicolumn{1}{c}{}&\multicolumn{2}{c}{$p=10$} & \multicolumn{2}{c}{$p=100$}& \multicolumn{2}{c}{$p=10$} & \multicolumn{2}{c}{$p=100$}\\
\cmidrule(l{3pt}r{3pt}){3-4}\cmidrule(l{3pt}r{3pt}){5-6}\cmidrule(l{3pt}r{3pt}){7-8}\cmidrule(l{3pt}r{3pt}){9-10}
Method& $\kappa$&$r=0.1$& $r=0.2$ & $r=0.1$& $r=0.2$ & $r=0.1$& $r=0.2$ & $r=0.1$& $r=0.2$ \\
\midrule
RMST-BART-default &0.25&2.61&3.32&3.27&3.90&1.51&2.12& 1.70&2.17\\
RMST-BART &0.25&1.52&2.01&2.07&2.77&0.73&1.07&1.14& 1.64\\
\midrule
RMST-BART-default &0.5&1.48&2.14&2.46&3.15&1.10&1.41&1.33&1.66\\
RMST-BART &0.5&1.54&2.10&2.37&2.82&0.62&0.83&1.08&1.64\\
\midrule  
RMST-BART-default &2&1.41&1.78&2.07&2.90&0.70& 0.90&1.05&1.43\\
RMST-BART &2&1.27&1.81&2.09&2.95&0.71&0.92&1.07& 1.49\\
\midrule  
RMST-BART-default &4&2.62&3.05&3.32&4.00&1.50&1.99&1.70&2.10\\
RMST-BART &4&2.91&3.56&3.74&4.53&1.51&2.16&1.74&2.41\\
\bottomrule
\end{tabular}
\caption{Friedman function simulation study with noninformative censoring. Comparing the average RMSE results of the RMST-BART-default and RMST-BART approaches across four different values of $\kappa$.}
\label{tab:friedmanKappa} 
\end{table}

Concluding the presentation of our simulation results, Table \ref{tab:linearAR1} compares the average RMSE between the RMST-BART method and alternative approaches on the absolute value linear model simulation study described in Section \ref{sec:Sim-AR1}. In these simulation scenarios, RMST-BART performs quite well across all scenarios but is not always the top performer. In particular, the regularized Cox proportional hazards method performs the best when $n=250$ and remains a competitive method when $n=1000$. For the large sample size settings, i.e., $n=1000$, RMST-BART is the top performing method when $r = 1.8$ while AFT-BART has somewhat better performance when $r = 0.8$. Even though the model used 
for simulating survival outcomes 
is not exactly a Cox proportional hazards model, the strong performance of the regularized Cox proportional hazards method likely
stems from the fact that the distribution of the survival times can be related directly to a linear function of the covariates,
and RMST-BART often has a more impressive relative performance when the survival distribution depends on a nonlinear function of the covariates.
Nevertheless, when the sample size is equal to one thousand, 
RMST-BART has enough data to improve upon the performance of a regularized Cox model procedure.
Coverage results for this simulation study can be found in Appendix B.

\begin{table}
\small
\centering
\begin{tabular}[h]{lllllllll}
\toprule
\multicolumn{1}{c}{\# of observations} &  \multicolumn{4}{c}{$n=250$} & \multicolumn{4}{c}{$n=1000$}\\
\cmidrule(l{3pt}r{3pt}){2-5}\cmidrule(l{3pt}r{3pt}){6-9}
\multicolumn{1}{c}{\# of predictors} & \multicolumn{2}{c}{$p=10$} & \multicolumn{2}{c}{$p=50$}& \multicolumn{2}{c}{$p=10$} & \multicolumn{2}{c}{$p=50$}\\
\cmidrule(l{3pt}r{3pt}){2-3}\cmidrule(l{3pt}r{3pt}){4-5}\cmidrule(l{3pt}r{3pt}){6-7}\cmidrule(l{3pt}r{3pt}){8-9}
Method& $r=0.8$& $r=1.8$ & $r=0.8$& $r=1.8$ & $r=0.8$& $r=1.8$ & $r=0.8$& $r=1.8$ \\
\midrule
 RMST-BART&0.65&0.70&0.73&0.72&0.59&0.59&0.65&0.68\\
 Coxph &0.68&0.72&0.98&0.85&0.65&0.66&0.69&0.71\\
 Regularized Coxph&0.65&0.67&0.65&0.67&0.65&0.65&0.65&0.65\\
 IPCW-boost&0.68&0.71&0.69&0.72&0.65&0.68&0.65&0.68\\
 AFT-Linear  & 0.91& 1.06&1.18&1.44&0.88&0.99&0.93&1.08\\
 AFT-BART&0.70&0.74&0.73&0.78&0.56&0.63&0.62&0.68\\
 \bottomrule
\end{tabular}
\caption{Absolute value linear model simulation study. Average RMSE results for RMST-BART and related approaches. Censoring is noninformative with censoring times generated from the $\textrm{Gamma}(2.2, r)$ distribution. Two values are used for $r$: $0.8$ and $1.8$, which result in approximately $10-20\%$ and $40-50\%$ censoring rates, respectively.}
\label{tab:linearAR1} 
\end{table}

\section{Application}\label{sec:Application}
This section demonstrates the use of our methodology with a real-world dataset of breast cancer outcomes assembled by the Molecular Taxonomy of Breast Cancer International Consortium (METABRIC). This dataset comprises 1904 primary fresh-frozen breast cancer specimens that were clinically annotated and collected from tumor banks located in the UK and Canada (\citet{pereira2016somatic}). This dataset includes patient-level demographic and clinical variables, gene expression measures, and mutation information on $173$ genes. For our analysis, we focused 
on 30 covariates of interest which included $10$ variables containing gene expression measures.
Among these $1904$ observations and $30$ covariates, there was a modest level of missingness in three of the covariates.
To address this and prepare our data for analysis using the RMST-BART procedure, we removed all observations that had
one or more missing covariate values which resulted in a final dataset of 1839 observations that we used for our analyses.

The survival outcome we targeted was overall survival measured in months. 
Median follow-up time among the $1839$ cases used in our analysis was quite long --- 116 months, and roughly
$42\%$ of these observations were censored at the time of the last follow up. 
For our analysis, we set the RMST-restriction point to $\tau = 300$ months because this is the $96^{th}$ percentile
of the estimated censoring distribution and $300$ months is a round number that has a direct interpretation as $25$ years.

\begin{figure}[ht]
    \centering
    \begin{subfigure}[t]{1\textwidth}
        \centering
        \includegraphics[height=0.40\textheight, width=0.45\textheight]{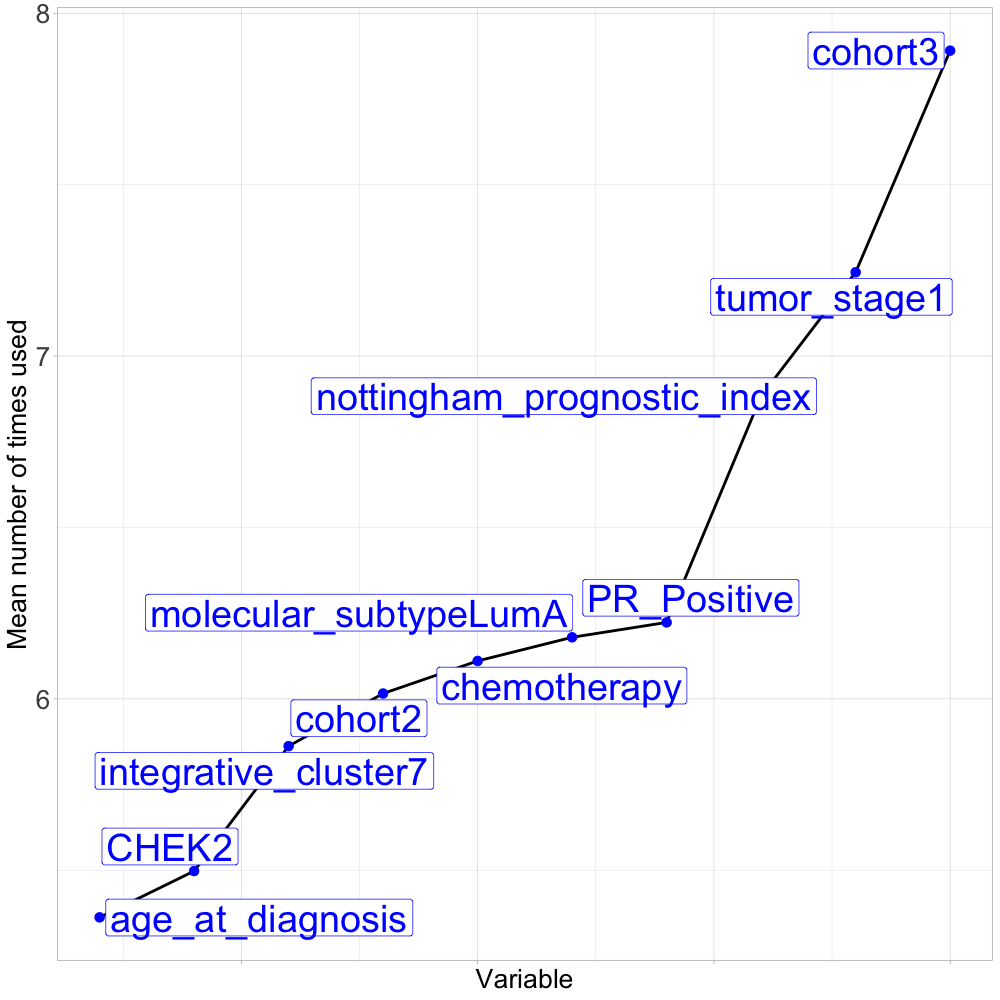} 
        \caption{Noninformative censoring} \label{fig:VarImp}
    \end{subfigure}
    \hfill
    \begin{subfigure}[t]{1\textwidth}
        \centering
        \includegraphics[height=0.40\textheight, width=0.45\textheight]{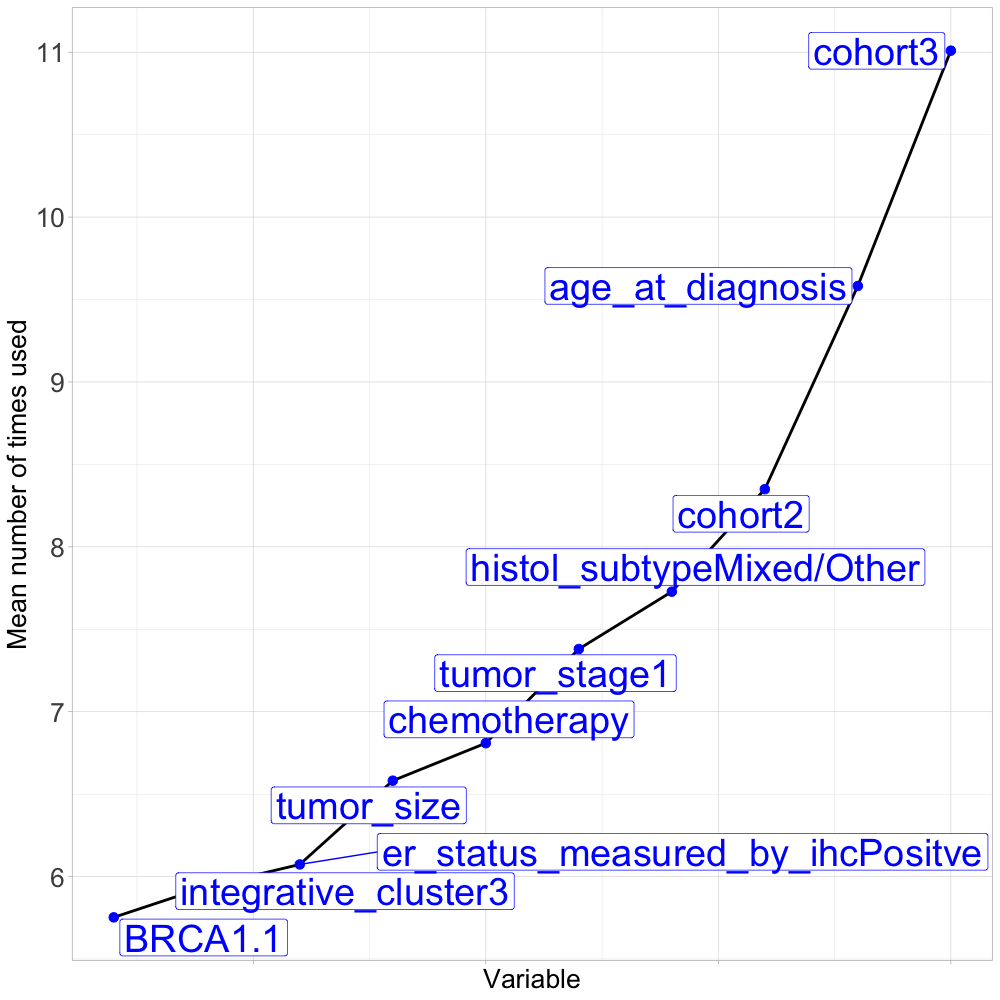} 
        \caption{Informative censoring} \label{fig:VarImpDep}
    \end{subfigure}
    \caption{ METABRIC data. Top 10 most frequently used variables by RMST-BART for both the noninformative and informative censoring models. In each case, RMST-BART used $200$ trees and $2000$ posterior draws. The y-axis represents the mean number of times a variable is used among the $200$ trees.}
\end{figure}

Figures \ref{fig:VarImp} and \ref{fig:VarImpDep} present RMST-BART variable importance plots of the top 10 variables from both the noninformative and informative censoring models. These variable importance measures, commonly used in other applications of BART, are averages (i.e., an average across posterior draws) of the number of times a variable is used as a splitting variable among the $200$ BART trees.
A comparison of these two plots reveals considerable disagreement in the top 10 variables between 
the informative and noninformative censoring models. 
Nevertheless, there are three variables that the informative and noninformative censoring models have in common: cohort membership, 
tumor stage, and patient age at the time of diagnosis. Regarding the cohort membership variable, this dataset divides the cases into five 
separate cohorts based on the time of sample collection, and the binary variable denoting membership in cohort 3 
was determined to be the most important variable by both the informative and noninformative censoring models of RMST-BART.
Membership in cohort 2 was also deemed to be one of the top 10 variables by both the informative and noninformative censoring models.
The ten gene expression variables tended to be somewhat less important according to this measure of RMST-BART variable importance, but 
expression levels for CHEK2 and BRCA1 were consistently the most important among the gene expression variables.

\begin{figure}[ht]
    \centering
    \begin{subfigure}[t]{1\textwidth}
        \centering
        \includegraphics[width=\linewidth]{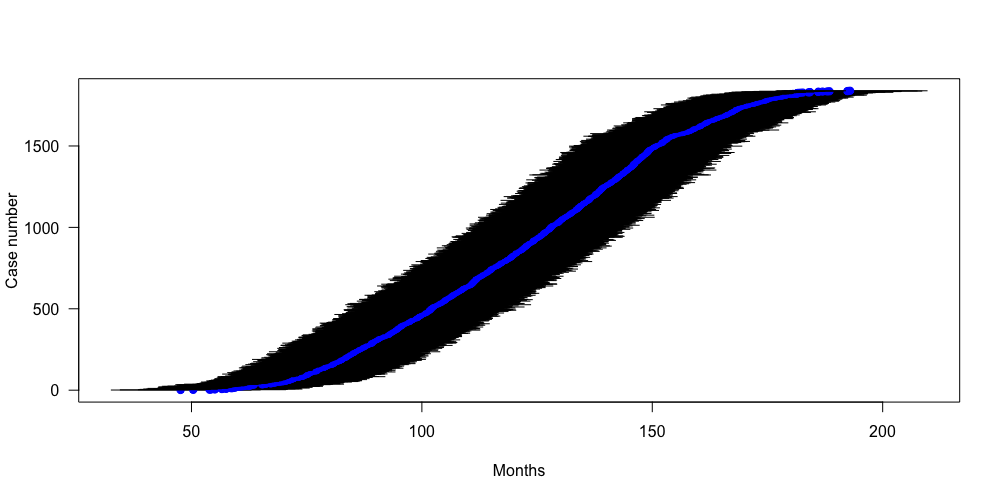} 
        \caption{Noninformative censoring} \label{fig:PIC}
    \end{subfigure}
    %\hfill
    \begin{subfigure}[t]{1\textwidth}
        \centering
        \includegraphics[width=\linewidth]{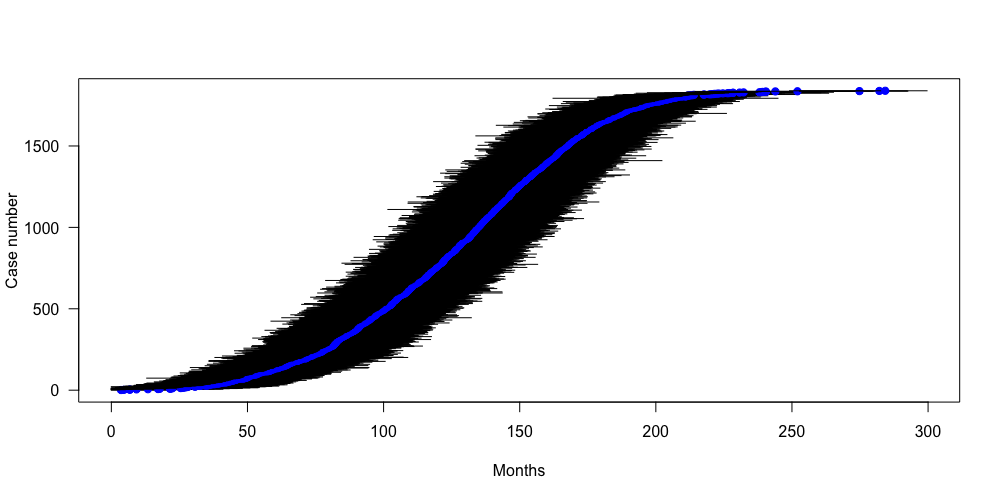} 
        \caption{Informative censoring} \label{fig:PICDep}
    \end{subfigure}
    \caption{METABRIC data. Test set posterior means and credible intervals using the RMST-BART algorithm with $200$ trees and $2000$ posterior draws. Blue points represent the point estimate for each case. Posterior means and credible intervals from the noninformative censoring model are shown in the top panel while posterior means and credible intervals from the informative censoring model are shown in the bottom panel.}
\end{figure}

The RMST-BART procedure can directly generate posterior means and credible intervals for $\mu_{\tau}(\mathbf{x}_{i})$
at each covariate vector $\mathbf{x}_{i}$ represented in the study.
Forest plots of posterior mean point estimates and $95\%$ uncertainty intervals for the $\mu_{\tau}(\mathbf{x}_{i})$ can provide
a visual summary of the variation in the values of $\mu_{\tau}(\mathbf{x}_{i})$ and the uncertainty
associated with the inference made on each individual point $\mu_{\tau}(\mathbf{x}_{i})$. 
Figures \ref{fig:PIC} and \ref{fig:PICDep} present the RMST-BART posterior means and $95\%$ credible intervals from both the informative and noninformative censoring models. 
When comparing these two figures, it is apparent that the individual-level credible intervals are consistently wider in the informative censoring model and that the spread of the posterior means of the $\mu_{\tau}(\mathbf{x}_{i})$ is notably greater in the informative censoring model when compared to the noninformative censoring model. These differences are largely due to both the greater posterior uncertainty associated with the individual-level values of the 
cumulative hazard $\Lambda(U_{i}^{\tau}|\mathbf{x}_{i})$ and the greater range of the posterior means of 
$\Lambda(U_{i}^{\tau}|\mathbf{x}_{i})$ in the informative censoring model.
Indeed, while both the informative and noninformative censoring models have a roughly similar median of $125$ months from their posterior mean point estimates of $\mu_{\tau}(\mathbf{x}_{i})$, the informative censoring model has a number of cases where the posterior mean is less than $50$ and a number of cases where the posterior mean of $\mu_{\tau}(\mathbf{x}_{i})$ exceeds $200$.
In contrast, essentially all of the posterior mean estimates in the noninformative censoring model are between $50$ and $200$.

For ensemble methods such as RMST-BART that are more ``black box'' in nature,
partial dependence plots (\citet{friedman2001}) can be a useful way to visually assess the impact of a particular covariate
has on the posterior mean of the RMST function $\hat{\mu}_{\tau}( \mathbf{x})$. The partial dependence function
for the $k^{th}$ covariate is defined as the expectation of a function estimator at a specific value
of covariate $k$, where the expectation is taken with respect to the joint distribution of all covariates except $k$.
In our context where we are interested in estimating the RMST function $\mu_{\tau}( \mathbf{x} )$,
the natural way of defining the estimated partial dependence function $\hat{\rho}_{k}( u )$ for covariate $k$ at point $u$ is 
\begin{equation}
    \hat{\rho}_{k}( u ) = \frac{1}{n}\sum_{i=1}^{n}\hat{\mu}_{\tau}(\mathbf{x}_{i,-k}, u),
\label{eq:pdfunction}
\end{equation}
where $\hat{\mu}_{\tau}(\mathbf{x}_{i,-k}, u)$ denotes the posterior mean of $\mu_{\tau}(\mathbf{x}_{i,-k}, u)$.
In (\ref{eq:pdfunction}), $\mathbf{x}_{i,-k}$ is the vector that contains all of the covariates for individual $i$ except 
for covariate $k$, and the notation $(\mathbf{x}_{i,-k}, u)$ denotes the vector where $x_{ik} = u$ 
and the remaining covariates are set to their observed values, i.e., 
$(\mathbf{x}_{i,-k}, u) = (x_{i1}, \ldots, x_{ik-1}, u, x_{ik+1}, \ldots, x_{ip})^{T}$.
Plotting $\hat{\rho}_{k}(u)$ versus $u$ can provide a useful summary of how changing $x_{ik}$ 
influences the estimated RMST.

\begin{figure}[ht]
		\centering
		\includegraphics[width=1\textwidth]{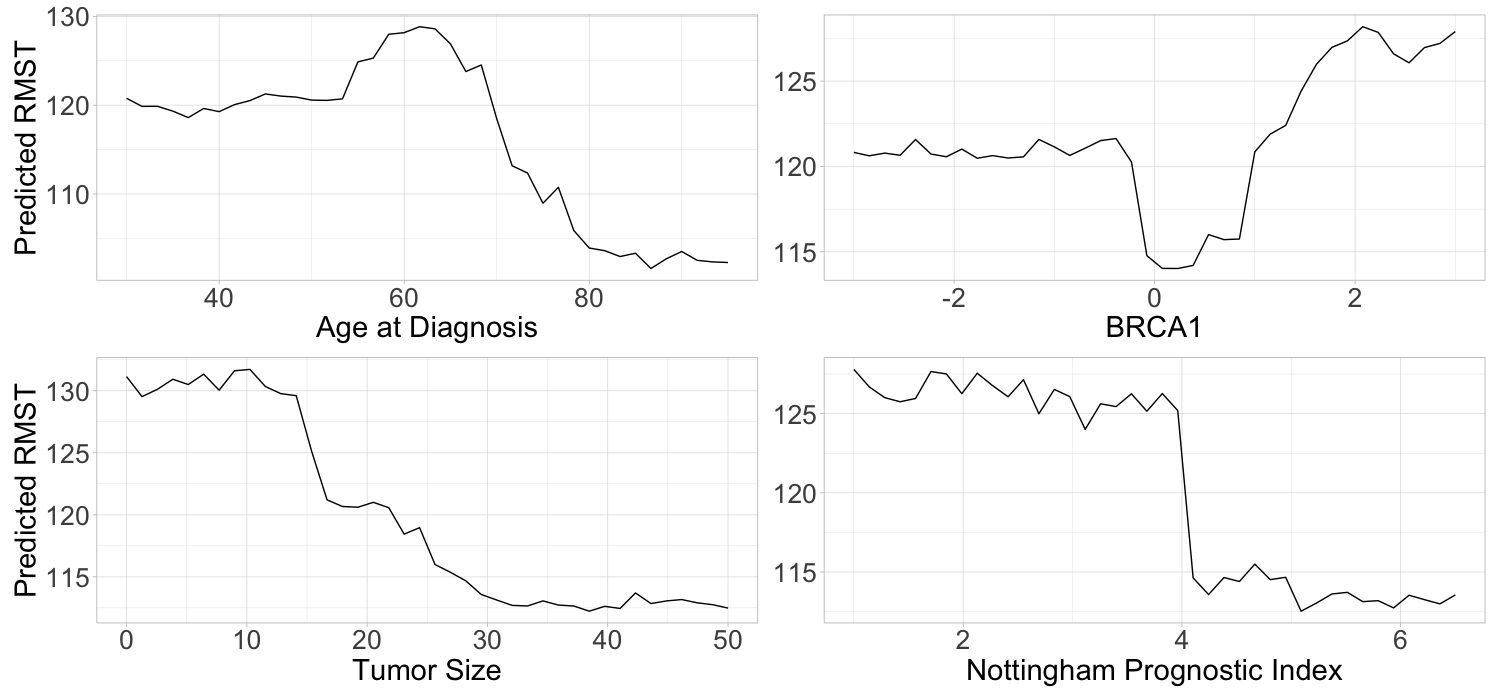}
             %\rotatebox[origin=c]{90}{\includegraphics[width=1.2\textwidth]{ParDep.png}}
		\caption{METABRIC data. RMST-BART partial dependence plots for the following key continuous variables: age at diagnosis, BRCA1 gene expression, tumor size, and the Nottingham Prognostic Index.}
		\label{fig:ParDep}
\end{figure}

Figure \ref{fig:ParDep} presents RMST partial dependence plots for 
four important continuous as identified by the RMST-BART variable importance scores.
These four covariates are patient age at diagnosis, expression measurements for the BRCA1 gene, tumor size, 
and the ``Nottingham Prognostic Index'' (\citet{galea1992}). 
From looking at the partial dependence plot for age at diagnosis, we can see that the highest estimated RMST values are observed in the age ranges of $55$ to $65$, followed by a clear downward trend after age 65. The steady decline after age 65 is
unsurprising, but, curiously this partial dependence plot suggests moderately worse survival outcomes in younger age groups when compared to the $50-55$ age range. While this could be driven by estimation variability or other factors, this observed trend could be reflecting the fact that younger breast cancer patients have been observed to more frequently display more aggressive disease compared to older patients (\citet{chen2016effect}). 
The importance of BRCA1 in breast cancer prognosis has been well recognized, and
the partial dependence function involving BRCA1 gene expression does indeed show considerable variability across values of this
variable though the relationship between standardized BRCA1 expression and predicted RMST does appear quite complex --- predicted RMST remains constant in the ranges $-3$ to $0$, experiences a notable drop around $0$, and then exhibits a subsequent increase thereafter. The effect of tumor size on the partial dependence function is quite expected with increasing tumor volume being monotonically associated with estimated RMST. 
The Nottingham Prognostic Index is an index where larger values of the index imply 
a worse prognosis for breast cancer patients, and the monotone decreasing partial dependence plot in Figure \ref{fig:ParDep} of this variable matches this interpretation. Interestingly, the partial dependence plot shows a sudden decrease in RMST as the Nottingham prognostic index moves from just below $4$ to just above $4$.
This quick decrease likely reflects the fact that the Nottingham Prognostic Index includes the sum of two 
categorical variables that are treated as numeric variables, and a move from one category to 
another in one of the important categorical components of the index
could certainly lead to a rapid decrease in the estimated RMST.

\begin{figure}[ht]
		\centering
		\includegraphics[width=1\textwidth]{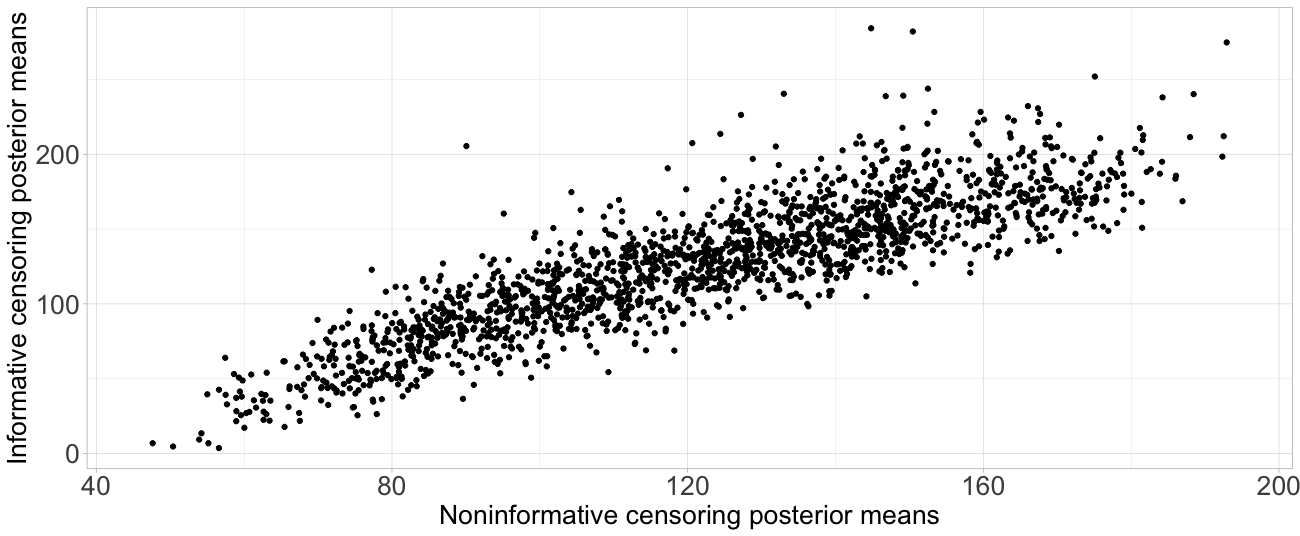}
		\caption{METABRIC data. RMST-BART posterior means under the informative censoring model versus RMST-BART posterior means under the noninformative censoring model.}
		\label{fig:PstComp}
\end{figure}

Figure \ref{fig:PstComp} provides a more direct, visual comparison of the RMST-BART posterior means of $\mu_{\tau}(\mathbf{x}_{i})$ 
generated under the informative censoring assumption versus the posterior means generated under the noninformative censoring model.
This figure shows a strong agreement between the posterior means of $\mu_{\tau}(\mathbf{x}_{i})$ produced under the two different assumptions about the censoring distribution. Indeed, the Pearson correlation between these two collections of posterior means was $0.87$. 
The most notable difference between these two sets of posterior means is the greater variability 
of the posterior means from the informative censoring model. 
%Indeed, for most of the points shown in this graph, 
%the noninformative censoring posterior mean can be viewed as a ``shrunken'' version of the informative 
%censoring posterior mean that is shrunken towards an overall mean of
%roughly ?.

\section{Discussion}\label{sec:Diss}
In this paper, we have introduced a framework for performing Bayesian inference on a function
which relates RMST to individual-level covariates. The motivation behind the development of this framework and method 
was to provide more robust Bayesian inference when the main inferential target of interest is the RMST values.
Our RMST-BART procedure addresses this motivating goal by utilizing a few key modeling strategies.
First, by using a generalized Bayes framework
focused on an RMST-targeted loss function, we have reduced the impact
of model misspecification of terms in a survival model that are not directly related to the RMST. 
Furthermore, by using a flexible BART prior for the RMST function, our approach can adapt to nonlinearities and covariate interactions in the RMST function and relies 
less on correctly specifying how the baseline covariates and RMST are connected.
Finally, by incorporating IPCW as a key component of the RMST-targeted loss function,
our approach can remove the need to employ complex joint modeling methods to handle informative censoring
thereby eliminating another source of modeling complexity and potential model misspecification.

RMST-BART is intended to be used in situations where one is mainly interested in performing 
inference on an RMST function that depends on a collection of baseline covariates. An advantage
of our generalized Bayes approach is that it does not require modeling features of the joint
distribution of $(T_{i}, C_{i})$ which are not relevant for estimating RMST. A drawback of 
adopting this approach is that one cannot generate a posterior distribution
over other features of the survival distribution of $T_{i}$ nor can one 
obtain a posterior predictive distribution that characterize the distribution
of future values of $T_{i}$, but if the primary goal is RMST inference,
this may often be a worthwhile tradeoff.

As we have outlined in this paper, a key feature of our procedure is the need to specify the tuning or ``learning rate'' parameter
$\eta$. We have explored two strategies for specifying this term --- one in which a AFT model is estimated to provide
a rough value of $\eta$ and one where cross-validation is used to further refine this preliminary value of $\eta$. 
While this strategy has shown good performance in our simulation studies, there can be a notable difference
in the value of $\eta$ chosen by the two different selection methods, particularly for the covariate-dependent censoring models. While the cross-validation approach typically has superior predictive performance for large sample sizes,
greater exploration of the relationship between the sample size, number of covariates,
and the performance of different selection methods for $\eta$ would be 
useful for informing the choice of $\eta$ when using RMST-BART in practice.
On a related point, we have observed in simulations that, 
in smaller sample settings, using the BART-based covariate-dependent model for censoring
often does not provide much of an advantage over a simple noninformative censoring model even
if the true generating model has informative censoring. Further examination
of the typical sample sizes needed for one to have a clear advantage from 
using the informative censoring model would be
helpful in practice when choosing the censoring model for RMST-BART.

Regarding the modeling of the censoring time distributions, we have assumed that one is interested
in only using the baseline covariates $\mathbf{x}_{i}$ in the model for the censoring probabilities. 
However, if one has additional time-varying covariates that one wants to include in a censoring model, these 
could easily be incorporated into our RMST inference framework as long as one can specify a tractable Bayesian model relating such time-dependent covariates to the censoring cumulative hazard. Indeed, if one can draw posterior 
samples for this cumulative hazard evaluated at the observed follow-up times, one can use these directly 
as the inverse probability of censoring weights in the RMST-BART procedure. This will not require modifying
our MCMC posterior sampling scheme as one can just use the draws of the inverse probability of censoring weights directly 
in all the other steps of the MCMC algorithm.

\medskip
\begin{center}
{\large\bf SUPPLEMENTARY MATERIAL}
\end{center}

We also have an R package called \verb"rmstbart" which implements the RMST-BART method and which is available for download at \url{https://github.com/nchenderson/rmstbart}.
Also, the datasets and R codes used in this paper are publicly available at 
\url{https://github.com/mahsaashouri/Loss-Function-BART-RMST}.

\appendix

\section{Conditional Distribution of $\bmu_{h}$}
The first thing to note is that
\begin{equation}
p( \bmu_{h}| \Lambda, \mathcal{T}_{h}, \mathcal{T}_{-h}, \bmu_{-h}, \mathbf{Y}^{\tau}, \bdelta)
\propto \exp\Big\{ -\eta \sum_{i=1}^{n} \delta_{i}\exp\{ \Lambda(U_{i}^{\tau}|\mathbf{x}_{i}) \}[ g_{h}(\mathbf{x}_{i}) - R_{ih} ]^{2} \Big\} \pi(\bmu_{h}|\mathcal{T}_{h} ),
\label{eq:cond_form}
\end{equation}
where $g_{-h}(\mathbf{x}_{i}) = \sum_{k\neq h} g_{k}(\mathbf{x}_{i};\mathcal{T}_{h}, \bmu_{h})$ and $R_{ih} = Y_{i}^{\tau} - g_{-h}(\mathbf{x}_{i})$.

Let $\mathbf{R}_{h}$ denote the $n \times 1$ vector of residuals $\mathbf{R}_{h} = (R_{1h}, \ldots, R_{nh})^{T}$.
Let $\mathbf{W}$ be the $n \times n$ diagonal matrix whose $i^{th}$ diagonal element is $\delta_{i}\exp\{\Lambda(U_{i}^{\tau}|\mathbf{x}_{i})\}$.
Let $\mathbf{A}_{h}$ be the $n \times n_{h}$ matrix (where $n_{h}$ is the length of $\bmu_{h}$)
whose $(i,j)$ entry is defined as
\begin{equation}
A_{ij,h} = 
\begin{cases} 
1 & \textrm{ if $\mathbf{x}_{i}$ is assigned to terminal node $j$ of tree $\mathcal{T}_{h}$} \nonumber \\
0 & \text{ otherwise } \nonumber
\end{cases}
\end{equation}
Note that $g_{h}(\mathbf{x}_{i}) = \mathbf{a}_{i,h}^{T}\bmu_{h}$, where $\mathbf{a}_{i,h}$ is the
$i^{th}$ row of $\mathbf{A}_{h}$.

Then, using the fact that $\pi(\bmu_{h}|\mathcal{T}_{h}) \propto \exp\{-\bmu_{h}^{T}\bmu_{h}/2\sigma_{\mu}^{2} \}$, we can 
re-write (\ref{eq:cond_form}) as 
\begin{eqnarray}
p( \bmu_{h}| \Lambda, \mathcal{T}_{h}, \mathcal{T}_{-h}, \bmu_{-h}, \mathbf{Y}^{\tau}, \bdelta) &\propto& \exp\Big\{ -\eta (\mathbf{R}_{h} - \mathbf{A}_{h}\bmu_{h})^{T}\mathbf{W}  (\mathbf{R}_{h} - \mathbf{A}_{h}\bmu_{h}) - \frac{1}{2\sigma_{\mu}^{2}}\bmu_{h}^{T}\bmu_{h} \Big\} \nonumber \\
&\propto& \exp\Big\{ -\eta \Big[ - 2\bmu_{h}^{T}\mathbf{A}_{h}^{T}\mathbf{W}\mathbf{R}_{h} + \bmu_{h}^{T}\mathbf{A}_{h}^{T}\mathbf{W}\mathbf{A}_{h}\bmu_{h}  + \frac{1}{2\eta\sigma_{\mu}^{2}}\bmu_{h}^{T}\bmu_{h} \Big] \Big\} \nonumber \\
&=& \exp\Big\{ -\frac{\eta}{2} \Big[  - 4\bmu_{h}^{T}\mathbf{A}_{h}^{T}\mathbf{W}\mathbf{R}_{h} + \bmu_{h}^{T}(2\mathbf{A}_{h}^{T}\mathbf{W}\mathbf{A}_{h} + (\sigma_{\mu}^{2}\eta)^{-1}\mathbf{I}_{n_{h}}) \bmu_{h} \Big] \Big\} \nonumber \\
&\propto& \exp\Big\{ -\frac{\eta}{2} \Big[ (\bmu_{h} - \mathbf{d}_{h})^{T}(2\mathbf{A}_{h}^{T}\mathbf{W}\mathbf{A}_{h} + (\sigma_{\mu}^{2}\eta)^{-1}\mathbf{I}_{n_{h}})(\bmu_{h} - \mathbf{d}_{h})  \Big] \Big\}, 
\label{eq:conjugate_form}
\end{eqnarray}
where
\begin{eqnarray}
\mathbf{d}_{h} &=& 2\Big( 2\mathbf{A}_{h}^{T}\mathbf{W}\mathbf{A}_{h} + (\sigma_{\mu}^{2}\eta)^{-1}\mathbf{I}_{n_{h}} \Big)^{-1}\mathbf{A}_{h}^{T}\mathbf{W}\mathbf{R}_{h} \nonumber \\
&=& \Big( \mathbf{A}_{h}^{T}\mathbf{W}\mathbf{A}_{h} + (2\sigma_{\mu}^{2}\eta)^{-1}\mathbf{I}_{n_{h}} \Big)^{-1}\mathbf{A}_{h}^{T}\mathbf{W}\mathbf{R}_{h} \nonumber \\
&=& \Big( \mathbf{C}_{h} + (2\sigma_{\mu}^{2}\eta)^{-1}\mathbf{I}_{n_{h}} \Big)^{-1}\mathbf{B}_{h}^{T}\mathbf{R}_{h}
\nonumber
\end{eqnarray}
In the definition of $\mathbf{d}_{h}$, $\mathbf{B}_{h}$ is the $n \times n_{h}$ matrix whose $(i,j)$ entry is
\begin{equation}
B_{ij,h}
=\begin{cases} 
\delta_{i}\exp\{\Lambda(U_{i}^{\tau}|\mathbf{x}_{i})\} & \textrm{ if $\mathbf{x}_{i}$ is assigned to terminal node $j$ of tree $\mathcal{T}_{h}$} \nonumber \\
0 & \text{ otherwise } \nonumber
\end{cases}
\end{equation}
and $\mathbf{C}_{h}$ is the $n_{h} \times n_{h}$ diagonal matrix whose $k^{th}$ diagonal entry is
\begin{equation}
C_{kk,h} = \sum_{i=1}^{n} \delta_{i}\exp\{\Lambda(U_{i}^{\tau}|\mathbf{x}_{i})\}A_{ik,h}. \nonumber 
\end{equation}

To summarize, (\ref{eq:conjugate_form}) implies that
\begin{equation}
\bmu_{h}| \Lambda, \mathcal{T}_{h}, \mathcal{T}_{-h}, \bmu_{-h}, \mathbf{Y}^{\tau}, \bdelta \sim \textrm{Normal}\Big( \mathbf{a}_{h}, \mathbf{D}_{h} \Big), \nonumber 
\end{equation}
where the $j^{th}$ element of $\mathbf{a}_{h}$ can be simplified to 
\begin{equation}
\sum_{i=1}^{n} \delta_{i}\exp\{\Lambda(U_{i}^{\tau}|\mathbf{x}_{i})\}A_{ij,h}R_{ih} \Big/ \Big( (2\eta\sigma_{\mu}^{2})^{-1} +  \sum_{i=1}^{n} \delta_{i}\exp\{\Lambda(U_{i}^{\tau}|\mathbf{x}_{i})\}A_{ij,h} \Big), \nonumber 
\end{equation}
and the $j^{th}$ diagonal element of $\mathbf{D}_{h}$ can be simplified to 
\begin{equation}
\frac{1}{\eta}\Big( (2\eta\sigma_{\mu}^{2})^{-1} +  \sum_{i=1}^{n} \delta_{i}\exp\{\Lambda(U_{i}^{\tau}|\mathbf{x}_{i})\}A_{ij,h} \Big)^{-1}. \nonumber
\end{equation}

\section{Additional Simulation Results}

Table \ref{tab:fixedGCOV} compares coverage of $95\%$ credible intervals from RMST-BART with a ``fixed weights" variation of 
RMST-BART where the same IPCW weights are used in each MCMC iteration rather than sampled in each MCMC iteration.
The fixed IPCW weights used are obtained by taking the posterior mean of the collection of posterior draws of the IPCW weights.
The results shown in Table \ref{tab:fixedGCOV} were obtained from running $100$ replications of the Friedman simulation
study with independent censoring. As shown in Table \ref{tab:fixedGCOV}, the ``fixed weights" versions
of RMST-BART and RMST-BART-default have much poorer coverage than RMST-BART and RMST-BART-default respectively.

Table \ref{tab:linCOV} illustrates coverage outcomes in the absolute value linear model simulation study for the following four methods: RMST-BART-default (default choice of $\eta$), RMST-BART (selection of $\eta$ through cross-validation), AFT-BCART-default, and AFT-BART. 
Coverage is evaluated for $95\%$ credible intervals of each method. 
In all presented scenarios, our suggested approach, whether RMST-BART-default or RMST-BART, consistently closer to $95\%$ coverage compared to the alternative methods, with minimal differences observed between RMST-BART-default and RMST-BART. 

\begin{table}
\footnotesize
\centering
\begin{tabular}[h]{lllllllll}
\toprule
\multicolumn{1}{c}{\# of observations} &  \multicolumn{4}{c}{$n=250$} & \multicolumn{4}{c}{$n=1000$}\\
\cmidrule(l{3pt}r{3pt}){2-5}\cmidrule(l{3pt}r{3pt}){6-9}
\multicolumn{1}{c}{\# of predictors} & \multicolumn{2}{c}{$p=10$} & \multicolumn{2}{c}{$p=100$}& \multicolumn{2}{c}{$p=10$} & \multicolumn{2}{c}{$p=100$}\\
\cmidrule(l{3pt}r{3pt}){2-3}\cmidrule(l{3pt}r{3pt}){4-5}\cmidrule(l{3pt}r{3pt}){6-7}\cmidrule(l{3pt}r{3pt}){8-9}
Method& $r=0.1$& $r=0.2$ & $r=0.1$& $r=0.2$ & $r=0.1$& $r=0.2$ & $r=0.1$& $r=0.2$ \\
\midrule
 RMST-BART & 0.92 & 0.87 & 0.88 & 0.78 & 0.94 & 0.93 & 0.94 & 0.88 \\
 RMST-BART-fixedG & 0.88 & 0.82 & 0.82 & 0.69 & 0.93 & 0.91 & 0.92 & 0.80 \\
  RMST-BART-default & 0.86 & 0.81 & 0.78 & 0.70 & 0.92 & 0.89 & 0.91 & 0.87 \\
 RMST-BART-default-fixedG & 0.73 & 0.67 & 0.61 & 0.53 & 0.88 & 0.80 & 0.86 & 0.79 \\
 \bottomrule
\end{tabular}
\caption{Coverage results for RMST-BART where a different set of IPCW weights are drawn
in each MCMC iteration (RMST-BART and RMST-BART-default) versus a version of RMST-BART where
the same sets of IPCW weights are used in each MCMC iteration (RMST-BART-fixedG and RMST-BART-default-fixedG).}
\label{tab:fixedGCOV} 
\end{table}

\begin{table}
\footnotesize
\centering
\begin{tabular}[h]{lllllllll}
\toprule
\multicolumn{1}{c}{\# of observations} &  \multicolumn{4}{c}{$n=250$} & \multicolumn{4}{c}{$n=1000$}\\
\cmidrule(l{3pt}r{3pt}){2-5}\cmidrule(l{3pt}r{3pt}){6-9}
\multicolumn{1}{c}{\# of predictors} & \multicolumn{2}{c}{$p=10$} & \multicolumn{2}{c}{$p=50$}& \multicolumn{2}{c}{$p=10$} & \multicolumn{2}{c}{$p=50$}\\
\cmidrule(l{3pt}r{3pt}){2-3}\cmidrule(l{3pt}r{3pt}){4-5}\cmidrule(l{3pt}r{3pt}){6-7}\cmidrule(l{3pt}r{3pt}){8-9}
Method& $r=0.8$& $r=1.8$ & $r=0.8$& $r=1.8$ & $r=0.8$& $r=1.8$ & $r=0.8$& $r=1.8$ \\
\midrule
 RMST-BART-default& 0.95 &0.86&0.95&0.88&0.95&0.87&0.94&0.85\\
% RMST-BCART-default &0.78&0.79&0.77&0.75&0.68&0.75&0.69&0.72\\
 AFT-BART-default &0.73&0.73&0.79&0.87&0.77&0.78&0.77&0.78\\
 RMST-BART &0.94&0.85&0.96&0.88&0.94&0.88&0.94&0.77\\
%RMST-BCART  &0.78&0.77&0.75&0.75&0.66&0.73&0.65&0.72\\
 AFT-BART &0.73&0.73&0.79&0.78&0.77&0.78&0.77&0.78\\
 \bottomrule
\end{tabular}
\caption{Coverage results comparisons for the RMST-BART, RMST-BCART, and AFT-BART methods for different settings
of $n$, $p$, and $r$ in the absolute value linear model simulation study.}
\label{tab:linCOV} 
\end{table}

\bibliographystyle{agsm}
\bibliography{RMSTBART_new}

\end{document}